\documentstyle[aps,prd,epsf,twocolumn]{revtex}

\newcommand{\be}{\begin{equation}}
\newcommand{\ee}{\end{equation}}
\newcommand{\bea}{\begin{eqnarray}}
\newcommand{\eea}{\end{eqnarray}}
\newcommand{\bdm}{\begin{displaymath}}
\newcommand{\edm}{\end{displaymath}}
\newcommand{\ul}{\underline}

\newcommand{\diff}{d}
\newcommand{\Diff}{D}

\newcommand{\p}{\partial}

\newcommand{\Trace}{\mbox{Tr}}
\newcommand{\sprod}[2]{\langle #1\, , \,#2 \rangle}

\newcommand{\identy}{1 \! \! 1}

\newcommand{\const}{\mbox{const.}}

\newcommand{\RR}{\mbox{$I \! \! R$}}

\newcommand{\bbb}{\mbox{\boldmath $b$}}

\newcommand{\bbB}{\mbox{\boldmath $B$}}

\newcommand{\bbL}{\mbox{\boldmath $L$}}
\newcommand{\bbP}{\mbox{\boldmath $P$}}

\newcommand{\bbS}{\mbox{\boldmath $S$}}

\newcommand{\gtens}{\mbox{\boldmath $g$}}
\newcommand{\Ktens}{\mbox{\boldmath $K$}}

\newcommand{\ghattens}{\hat{\gtens}}
\newcommand{\gbartens}{\mbox{\boldmath $\bar{g}$}}

\newcommand{\ghat}{\hat{g}}
\newcommand{\gbar}{\bar{g}}
\newcommand{\Abar}{\bar{A}}

\newcommand{\nabhat}{\hat{\nabla}}
\newcommand{\asthat}{\hat{\ast}}

\newcommand{\lapbar}{\bar{\Delta}}
\newcommand{\nabbar}{\bar{\nabla}}
\newcommand{\astbar}{\bar{\ast}}
\newcommand{\etabar}{\bar{\eta}}
\newcommand{\Diffbar}{\bar{\Diff}}

\newcommand{\chris}[2]{\Gamma^{#1}_{\; #2}}
\newcommand{\chrishat}[2]{\hat{\Gamma}^{#1}_{\; #2}}

\newcommand{\chrisbar}[2]{\bar{\Gamma}^{#1}_{\; #2}}

\begin{document}
\twocolumn[\hsize\textwidth\columnwidth\hsize\csname @twocolumnfalse\endcsname

\title{Self-adjoint wave equations for dynamical perturbations of
       self-gravitating fields}

\author{O. Sarbach$^{\star}$, M. Heusler$^{\star \dagger}$, and O. Brodbeck$^{\dagger}$}
\address{$^{\star}$Institute for Theoretical Physics, University of Zurich,
         CH--8057 Zurich, Switzerland \\ Time-steps GMbH, CH-8910 Zurich, Switzerland}
\date{\today}

\maketitle

\begin{abstract}
It is shown that the dynamical evolution of linear perturbations
on a static space-time is governed by a constrained wave equation
for the extrinsic curvature tensor. The spatial part of the wave
operator is manifestly elliptic and self-adjoint.
In contrast to metric formulations, the curvature-based approach to
gravitational perturbation theory generalizes in a natural way to 
self-gravitating matter fields.
It is also demonstrated how to obtain symmetric pulsation equations
for self-gravitating non-Abelian gauge fields, Higgs fields and
perfect fluids.
For vacuum fluctuations on a vacuum space-time, the Regge-Wheeler
and Zerilli equations are rederived.\\
\end{abstract}
]

\section{Introduction}

Perturbation theory finds several interesting applications
in general relativity. A prominent example is the close limit
approximation to black hole collisions \cite{PP94}.
Besides providing estimates for the energy emitted by gravitational radiation,
the close limit results also play an important role in testing
existing numerical codes for full non-linear general relativity.
Further examples comprise the linear stability analysis of
neutron stars or black holes with matter fields.
Recently, a perturbation approach has also been used to find new stationary 
solutions and to prove local uniqueness results for hairy black holes and 
self-gravitating solitons \cite{BHSV} \cite{SHB-Odd} \cite{SHB-Even}.

For vacuum gravity perturbation theory is highly developed,
and even for self-gravitating systems important general
properties of the perturbation equations were established.
In particular, it has been shown that a manifestly hyperbolic,
gauge-invariant formulation of gravitational perturbation theory
does exist for static background configurations \cite{CBYA-Hyperbolic},
\cite{AAL-CurvPert}.
On the other hand, it is known for a wide class of matter models
that the pulsation equations governing spherical perturbations
of static, spherically symmetric field configurations are
manifestly symmetric and derivable from the second variation of
the Komar mass \cite{BMS-Pulsation}.
However, only for special systems, such as vacuum gravity
and the Einstein-Maxwell system for a spherically symmetric
background, formulations are known for which the perturbations are governed
by a wave equation which is hyperbolic {\it and\/} symmetric. 
Symmetric formulations are most valuable, since they provide the possibility
to discuss stability issues by rigorous means, applying methods from spectral theory. 

The commonly adopted procedure to derive the perturbation equations
for spherically symmetric black holes is to first use the
symmetry of the background in order to expand the metric in terms of
spherical harmonics. Then, the pulsation equations are derived for 
gauge-invariant quantities. For vacuum perturbations
of vacuum gravity, this leads to the Regge-Wheeler equation
\cite{RW} in the odd-parity sector and to the Zerilli equation
\cite{Zerilli} for even-parity perturbations. Both equations have the
form of symmetric wave equations and, as an important consequence,
the stability of the Schwarzschild metric within linear perturbation theory 
can be established.
This result was extended to the Einstein-Maxwell system by Moncrief
\cite{M-RN}. While also starting with an expansion of the metric in spherical
harmonics, Moncrief uses the ADM Hamiltonian to find
a symmetric wave equation. In particular, he introduces gauge-invariant
quantities and at the same time, separates the constraint
variables from the dynamical variables.

As we have argued in a recent letter \cite{BHS-Letter}, the metric
approach fails to yield hyperbolic and symmetric equations
in the presence of general gravitating matter fields:
In order to obtain a wave operator appearing already
off-shell, one has to introduce amplitudes which are adapted to the
{\it stationarity\/} rather than the spherical symmetry of the background.
Using ideas introduced by Choquet-Bruhat {\it et al.\/} for full,
non-linear general relativity \cite{CBYA-Hyperbolic}, we show that a
hyperbolic, symmetric wave equation for the {\it extrinsic curvature\/}
can be obtained by natural means when the background is {\it static\/}.
In contrast to the traditional, {\it metric\/} based approach,
our {\it curvature\/} based approach is manifestly gauge-invariant,
hyperbolic and symmetric without using the spherical symmetry
of the background.
More importantly, we also show that a natural generalization to
self-gravitating fields is possible, including non-Abelian gauge
fields and perfect fluids.

This work is organized as follows:
In Section II, we show how to obtain a manifestly hyperbolic and symmetric
wave operator for the perturbation of the extrinsic curvature on a
static background. Then, in Section III, we specialize the result on a vacuum
background and discuss the initial value formulation and the projection
onto the constraint manifold. 
The coupling to Yang-Mills-Higgs fields and to perfect fluids
is discussed in Sections IV and V, respectively.
In Appendix A, we show how to separate the constraint and dynamical
variables and re\-derive the Regge-Wheeler and Zerilli equations
in quite a natural way.
Finally, in Appendix B, we recall a useful field theoretical formulation
of perfect fluids and show how this formulation applies to perturbation
theory.

\section{The construction of the wave operator}

The goal is to linearize the equations governing self-gravitating
matter fields for a {\it static\/} background and to bring the
perturbation equations into the form of a {\it symmetric wave equation\/},
\be
\left( \p_t^{\, 2} + {\cal A} \right) u = 0,
\label{Eq-SymWaveEq}
\ee
where $u$ describes the perturbed gravitational and matter fields,
and where ${\cal A}$ is a (formally) self-adjoint, elliptic
operator containing spatial derivatives up to second order.
In particular, this implies that equation (\ref{Eq-SymWaveEq})
is {\it hyperbolic\/}, which guarantees a well-posed initial
value formulation.
Furthermore, as the spatial operator ${\cal A}$ is self-adjoint,
methods form spectral theory can be applied and stability issues can 
be discussed analytically.

In order to give some motivation for the ideas developed below, 
we construct the wave operator belonging to the source-free Maxwell 
equations in flat space-time.
In view of a generalization to non-Abelian gauge groups it
is convenient to formulate Maxwell's equations in terms of
the electric one-form, $E$, and the magnetic potential one-form,
$A$, where $B = \ast\diff A$. 
Maxwell's equations constitute a system of constrained
evolution equations, with Gauss constraint
\bdm
\diff^\dagger E = 0,
\edm
and evolution equations
\bdm
\dot{A} = \diff\phi - E, \;\;\;
\dot{E} = \diff^\dagger\diff A.
\edm
Here, the codifferential operator $\diff^\dagger$ for a $p$-form $\omega$
is defined by $\diff^\dagger\omega \equiv (-1)^p \ast\diff\ast\omega$. 
Note that there is no evolution equation for the electric
scalar potential $\phi$. This reflects the gauge freedom of the theory.
Usually, a wave equation for $A$ is derived by imposing the
Lorentz gauge condition $\dot{\phi} + \diff^\dagger A = 0$.
However, $A$ is not gauge-invariant.
In order to obtain a hyperbolic wave equation in terms of the
gauge-invariant quantity $E$, we differentiate the second
evolution equation with respect to $t$, eliminate $\dot{A}$
using the first evolution equation and add the differential
of the constraint equation. This yields
\bdm
\ddot{E} + \left(\diff^\dagger\diff + \diff\,\diff^\dagger\right) E = 0,
\edm
which is of the desired form (\ref{Eq-SymWaveEq}).
The initial value problem may be solved as follows:
First, any function $\phi(t,x)$ is chosen.
Next, initial data $A_{t=0}$ and  $E_{t=0}$ are given, where
$E_{t=0}$ is subject to the Gauss constraint. Then,
$\dot{E}_{t=0}$ is computed from the second evolution equation, 
where $\diff^\dagger\dot{E}_{t=0} = 0$ automatically follows.
In a next step, $E_t$ is computed for all times using the symmetric
wave equation.
Finally, the magnetic potential is obtained from
\bdm
A_t = A_{t=0} + \int\limits_0^t ( \diff\phi_\tau - E_\tau)\,\diff\tau.
\edm
The Gauss constraint and the second evolution
equation, which was differentiated in order to get the
wave operator, are both constraint equations for the initial
data. It is easy to see that they propagate.

A gauge-invariant wave operator governing linear fluctuations of
self-gravitating fields on a static background
can be constructed in a similar manner.
First, we notice that a manifestly symmetric formulation of the
perturbation equations for a vacuum space-time exist:
\be
\frac{ \delta\left( \sqrt{-g}\, G^{\mu\nu} \right) }{\sqrt{-g}} =
\frac{1}{2} \eta^{\mu\alpha\sigma\tau} \eta^{\nu\beta\rho}_{\quad\;\tau}
\nabla_{(\alpha} \nabla_{\beta)} \delta g_{\sigma\rho} + G^{\alpha\beta\mu\nu} \delta g_{\alpha\beta}.
\label{Eq-DeltaGmunuSym}
\ee
Here, $G_{\mu\nu}$ is the Einstein tensor, $\eta$ is the volume form on $M$ and
the tensor $G^{\alpha\beta\mu\nu}$ is given in terms of the Riemann tensor,
$R^\alpha_{\;\beta\mu\nu}$, the Ricci tensor, $R_{\beta\nu} = R^\mu_{\;\beta\mu\nu}$,
and the Ricci scalar, $R = R^\mu_{\;\mu}$, by
\bea
&& G^{\alpha\beta\mu\nu} = -\frac{1}{2} R^{\alpha\mu\beta\nu} \nonumber\\
&& + \frac{1}{4} \left( 2g^{\alpha\beta} g^{\mu\nu} + 2g^{\mu\nu} R^{\alpha\beta}
  - 3g^{\alpha\mu} R^{\beta\nu} - 3g^{\beta\nu} R^{\alpha\mu} \right) \nonumber\\
&& + \frac{1}{4} \left( g^{\mu\alpha} g^{\nu\beta} + g^{\mu\beta} g^{\nu\alpha}
  - g^{\alpha\beta} g^{\mu\nu} \right)R. \nonumber
\eea
Clearly, the operator on the right-hand side (RHS) of Eq. (\ref{Eq-DeltaGmunuSym})
is formally self-adjoint with respect to the inner product
\be
\sprod{\delta g^{(1)}}{\delta g^{(2)}} = \int 
 g^{\alpha\mu} g^{\beta\nu} \delta g^{(1)}_{\alpha\beta}\,\delta g^{(2)}_{\mu\nu}\,\eta.
\label{Eq-InnerProduct}
\ee
In order to get an evolution equation for the perturbed geometry
on a static background, our program is the following:
\begin{enumerate}
\item{First, we perform a 3+1 decomposition of Eq. (\ref{Eq-DeltaGmunuSym}) for a
static background. The resulting equations are, of course, still
symmetric with respect to the inner product (\ref{Eq-InnerProduct}), but split into
two sets, comprising the constraint equations and the evolution equations, respectively}
\item{Among the $10$ components of $\delta g_{\alpha\beta}\,$, $4$
correspond to infinitesimal coordinate transformations. Hence, one has to
fix this gauge, or better, to construct $6$ gauge-invariant amplitudes.
For a static background the components of
the linearized extrinsic curvature tensor are gauge-invariant,
up to a reparametrization of time. If, in addition, space-time
is spherically symmetric, full gauge-invariant quantities
can be constructed from the components of the extrinsic curvature tensor.}
\item{The perturbation equations will be put into the 
desired form (\ref{Eq-SymWaveEq}), where the spatial operator
${\cal A}$ is first required to be elliptic. In order to
achieve this, ideas introduced by Choquet-Bruhat {\it et. al.\/} \cite{CBYA-Hyperbolic}
are used. Extending their method, the spatial operator is finally made symmetric with respect
to the inner product (\ref{Eq-InnerProduct}), where
$g^{\alpha\mu} g^{\beta\nu}$ is replaced by its spatial part $\gbar^{ik}\gbar^{jl}$.}
\end{enumerate}

\subsection{The ADM equations}
\label{Sect-ADMEq}

The Arnowitt-Deser-Misner (ADM) formalism provides a $3+1$ 
decomposition of (the portion of) space-time with 
topology $M \equiv \RR\times \Sigma$, where
$\Sigma$ is a three-dimensional Riemannian manifold.
The manifold $M$ is foliated by a one-parameter family of embeddings
$e_t: \Sigma\rightarrow M$, $t\in\RR$, where the hypersurfaces
$\Sigma_t \equiv e_t(\Sigma)$ are assumed to be space-like.
With respect to this foliation, the timelike vector field 
$\p_t$ is decomposed according to $\p_t = \alpha\, n + \beta$,
where $n$ is a normal unit vector field orthogonal to
$\Sigma_t$, and $\beta$ is tangential to $\Sigma_t$.
The metric assumes the form
\be
\gtens = -\alpha^2 \diff t^2 + \gbar_{ij} (\diff x^i + \beta^i \diff t)(\diff x^j + \beta^j \diff t),
\label{Eq-ADMMetric}
\ee
where the  $x^i$ are local coordinates on $\Sigma$,
and $\gbartens$ is the induced Riemannian metric on $\Sigma_t$.
With respect to the metric (\ref{Eq-ADMMetric}), the Einstein tensor
becomes
\bea
G_{00} &=& \frac{1}{2} \left( \bar{R} - \bar{G}^{ijkl} K_{ij} K_{kl} \right), \label{Eq-ADMConsHam}\\
G_{i0} &=& \bar{G}_{ijkl} \nabbar^j K^{kl}, \label{Eq-ADMConsMom}
\eea
and
\bea
G_{ij} &=& \bar{G}_{ij} - 2 K^s_{\; i} K_{sj} + K K_{ij} + \frac{1}{2}\gbar_{ij} \left( 3 K^{rs} K_{rs} - K^2 \right)
        \nonumber\\
       &-& \frac{1}{\alpha} \bar{G}_{ij}^{\;\;\; kl} \nabbar_k\nabbar_l\alpha + \frac{1}{\alpha} \bar{G}_{ij}^{\;\;\; kl} (\p_t - \bar{L}_{\beta} ) K_{kl},
\label{Eq-ADMEvol}
\eea
where the index zero refers to the normal vector field
$e_0 = n = \frac{1}{\alpha} (\p_t - \beta)$.
The $K_{ij}$ denote the components of the extrinsic curvature,
\be
K_{ij} = \frac{1}{2\alpha} (\p_t - \bar{L}_{\beta} )g_{ij} = 
\frac{1}{2\alpha} \left( \p_t\gbar_{ij} - \nabbar_i\beta_j -\nabbar_j\beta_i \right),
\label{Eq-ADMExtrinsic}
\ee
where $K$ is the trace of $K_{ij}$, $K = \gbar^{ij} K_{ij}$.
All quantities with a bar refer to the Riemannian metric $\gbartens$.
The tensor $\bar{G}^{ijkl}$ is the De Witt metric on the space
of symmetric, positive definite matrices,
\bdm
\bar{G}^{ijkl} = \frac{1}{2}\left( \gbar^{ik} \gbar^{jl} + \gbar^{il} \gbar^{jk} 
  - 2\gbar^{ij}\gbar^{kl} \right).
\edm
Equations (\ref{Eq-ADMEvol}) and (\ref{Eq-ADMExtrinsic}) are a set
of evolution equations for the $3$-dimensional metric $\gbartens$ and the extrinsic
curvature $\Ktens$, whereas eqs. (\ref{Eq-ADMConsHam}) and
(\ref{Eq-ADMConsMom}) are {\it constraint\/} equations.
The freedom to choose the slicing is reflected by the fact that there
are no evolution equations for the lapse, $\alpha$,
and the shift, $\beta$. The Bianchi identities,
\be
0 = \nabla_{\mu} G^{\mu\nu} = \p_0 G^{0\nu} + \p_k G^{k\nu} + 
\chris{\mu}{\mu\sigma} G^{\sigma\nu} + \chris{\nu}{\mu\sigma} G^{\mu\sigma},
\label{Eq-ADMBianchi}
\ee
guarantee that the Hamiltonian and the momentum constraints, eqs. (\ref{Eq-ADMConsHam})
and (\ref{Eq-ADMConsMom}) respectively, propagate.

A convenient choice for the lapse and the shift
is given by the gauge conditions
\bdm
K = \const, \;\;\;
\beta = 0,
\edm
for which the ADM equations simplify considerably.
(For asymptotically flat space-times this implies the {\it maximal slicing\/}
condition $K=0$.) In particular, one has for constant $K$      
\be
\frac{\alpha}{2}\left( \gbar^{ij} G_{ij} - 3G_{00} \right)
= \left( \lapbar - \bar{R} - K^2 \right)\alpha.
\label{Eq-ADMalphaCons}
\ee
In vacuum, this yields an elliptic equation for the lapse $\alpha$,
which, in a numerical evolution scheme, has to be solved after each time step. 

\subsection{Coordinate-invariant quantities}
\label{Sect-ADMCoordInv}

In a static space-time it is convenient to choose an adapted
slicing (i.e. orthogonal to $\p_t$), such that $\beta = 0$,
$\p_t\alpha = 0$ and $\p_t\gbar_{ij} = 0$.
As a consequence, the extrinsic curvature tensor vanishes,
\bdm
K_{ij} = 0.
\edm
The metric then becomes $\gtens = -\alpha^2 \diff t^2 + \gbartens$,
while the ADM equations reduce to
\be
R_{00} = \frac{1}{\alpha}\lapbar\alpha, \;\;\;
R_{0j} = 0, \;\;\;
R_{ij} = \bar{R}_{ij} - \frac{1}{\alpha}\nabbar_i\nabbar_j\alpha.
\label{Eq-ADMBack}
\ee

Now consider a tensor field $T_{\mu\nu}$ on $M$, which is static
on the background, $\dot{T}_{\mu\nu} = 0$, $T_{tj} = T_{it} = 0$.
Under an infinitesimal coordinate transformation,
$x^\mu \mapsto x^\mu + \delta x^\mu$, generated by the vector
field $X^\mu = \delta x^\mu$, $\delta T_{\mu\nu}$ transforms
according to $\delta T_{\mu\nu} \mapsto \delta T_{\mu\nu} + L_X T_{\mu\nu}$. 
Writing $(X^\mu) = (X^t,X^i) = (f,X^i)$, one has
\bea
\delta T_{tt} &\mapsto& \delta T_{tt} + X^i T_{tt,i} + 2\dot{f}\, T_{tt}, \nonumber\\
\delta T_{tj} &\mapsto& \delta T_{tj} + \dot{X}^i T_{ij} + f_j T_{tt}, \\
\delta T_{ij} &\mapsto& \delta T_{ij} + X^s \nabbar_s T_{ij} + T_{is} \nabbar_j X^s + T_{sj} \nabbar_i X^s,
\nonumber
\eea
where here and in the following a dot denotes differentiation with respect to $t$,
and $f_j \equiv \p_j f$.
Using $\delta g_{tt} = -2\alpha\delta\alpha$ and $\delta g_{tj} = \delta\beta_j$,
as well as the formulae
$\alpha^2\delta T_{00} = \delta T_{tt} - 2 T_{tt}\delta\alpha/\alpha$,
$\alpha\delta T_{0j} = \delta T_{tj} - T_{ij} \delta\beta^i$,
the metric and the Einstein tensor transform according to
\bea
\delta\alpha &\mapsto& \delta\alpha + \alpha_i X^i + \alpha\dot{f}, \nonumber\\
\delta\beta_j &\mapsto& \delta\beta_j + \dot{X}_j - \alpha^2 f_j\, , \label{Eq-ADMCoordTrans}\\
\delta\gbar_{ij} &\mapsto& \delta\gbar_{ij} + \nabbar_i X_j + \nabbar_j X_i\, , \nonumber
\eea
and
\bea
\delta G_{00} &\mapsto& \delta G_{00} + G_{00,i}\, X^i, \nonumber\\
\delta G_{0j} &\mapsto& \delta G_{0j} + \alpha\left( G_{00} f_j + G_{ij} f^i \right), \\
\delta G_{ij} &\mapsto& \delta G_{ij} + X^s \nabbar_s G_{ij} + G_{is} \nabbar_j X^s + G_{sj} \nabbar_i X^s. \nonumber
\eea

The above transformation properties imply that one can construct ``vector-invariant''
quantities, i.e., quantities which are invariant under the subset of transformations
generated by vector fields $(X^\mu) = (f, X^i)$ with $f=0$:
\bea
\delta\dot{G}^{(inv)}_{00} &\equiv& \delta\dot{G}_{00} - G_{00,i} \delta\beta^i, \nonumber\\
\delta G^{(inv)}_{0j} &\equiv& \delta G_{0j}, \nonumber\\
\delta\dot{G}^{(inv)}_{ij} &\equiv& \delta\dot{G}_{ij} - \delta\beta^s\nabbar_s G_{ij} - G_{si}\nabbar_j\delta\beta^s - G_{sj}\nabbar_i\delta\beta^s, \nonumber
\eea
and
\bea
\delta\dot{\alpha}^{(inv)} &\equiv& \delta\dot{\alpha} - \alpha^j \delta\beta_j\, ,\nonumber\\
\delta\dot{\gbar}^{(inv)}_{ij} &\equiv& \delta\dot{\gbar}_{ij}
 - \nabbar_i\delta\beta_j - \nabbar_j\delta\beta_i. \label{Eq-ADMgijinv}
\eea
The fact that $\delta\dot{\gbar}^{(inv)}_{ij} = 2\alpha\delta K_{ij}$ suggests
that -- from the ADM point of view -- the natural gauge-invariant perturbations 
on a static background are the {\it extrinsic curvatures}, rather than the
metric fields. We are, therefore, looking for a symmetric wave equation
in terms of the perturbed extrinsic curvature tensor $\delta K_{ij}$.

\subsection{The linearized ADM equation}
\label{Sect-LinADMEq}

The derivation of the linearized field equations is considerably simplified by 
choosing an appropriate gauge. For a static background the vector field $X^j$ in Eq.
(\ref{Eq-ADMCoordTrans}) can be arranged such that $\delta\beta^j = 0$. (Note,
however, that this gauge is not adapted to describe stationary perturbations.)
In this gauge, the perturbed quantities $\delta\dot{\alpha}$,
$\delta\dot{\gbartens}$ and $\delta\dot{G}_{\mu\nu}$ coincide with
their vector-invariant counterparts constructed above.
The remaining residual gauge freedom is generated by the vector
field $(X^\mu) = (f, X^i)$ with $\dot{X}_j = \alpha^2 f_j$,
\bea
\delta\dot{\alpha} &\mapsto& \delta\dot{\alpha} + \alpha^2\alpha_j f^j + \alpha\ddot{f}, \nonumber\\
\delta K_{ij} &\mapsto& \delta K_{ij} + \frac{1}{\alpha} \nabbar_{(i} \left( \alpha^2\nabbar_{j)} f \right).
\label{Eq-ADMResGaugeTrans}
\eea
Here and in the following, we use the notation
$2\omega_{(ij)} \equiv \omega_{ij} + \omega_{ji}$.

Using the fact that quadratic terms in $\Ktens$ give no contributions if the background 
is static, the linearized ADM constraint equations become
\be
\delta G_{00} = \frac{1}{2} \delta\bar{R}, \;\;\;
\delta G_{i0} = \bar{G}_{ijkl} \nabbar^j \delta K^{kl},
\label{Eq-DeltaADMCons}
\ee
while the evolution equations are
\bea
\delta K_{ij} &=& \frac{1}{2\alpha} \delta\dot{\gbar}_{ij}, \label{Eq-DeltaADM1}\\
\delta G_{ij} &=& \delta\bar{G}_{ij} - \delta\,\frac{1}{\alpha} \bar{G}_{ij}^{\;\;\; kl} \nabbar_k\nabbar_l\alpha
               + \frac{1}{\alpha} \bar{G}_{ij}^{\;\;\; kl} \delta\dot{K}_{kl}. \label{Eq-DeltaADM2}
\eea
A wave equation for $\delta\Ktens$ is obtained as follows:
One first differentiates Eq. (\ref{Eq-DeltaADM2}) with respect to $t$
and uses Eq. (\ref{Eq-DeltaADM1}) to eliminate $\delta\dot{\gbartens}$.
Applying the general formulae 
\bea
2\delta R_{\alpha\beta} &=& -\nabla^\mu\nabla_\mu\,\delta g_{\alpha\beta} - \nabla_\alpha\nabla_\beta (g^{\mu\nu}\delta g_{\mu\nu}) \nonumber\\
&+& \nabla^\mu\nabla_\beta\,\delta g_{\alpha\mu} +  \nabla^\mu\nabla_\alpha\,\delta g_{\beta\mu} \nonumber
\eea
and
\bdm
\delta G_{\alpha\beta} = \delta R_{\alpha\beta} - \frac{1}{2} R\,\delta g_{\alpha\beta}
 - \frac{1}{2} g_{\alpha\beta} \left( g^{\mu\nu} \delta R_{\mu\nu} - R^{\mu\nu} \delta g_{\mu\nu} \right)
\edm
to the variation of the Einstein tensor $\bar{G}_{ij}$ in three dimensions,
and also using the background equations (\ref{Eq-ADMBack}), then yields
the following expressions in terms of the vector-invariant amplitudes:
\be
L_{ij} \equiv \alpha\delta K_{ij} = \frac{1}{2} \delta\dot{g}_{ij}, \;\;\;
A \equiv \delta\dot{\alpha},
\ee
\bea
S_{ij} &\equiv& \frac{\alpha}{\sqrt{-g}}\,\p_t\, \delta\left( \sqrt{-g}\, G_{ij} \right) \nonumber\\
 &=& S^{BG}_{ij} + \bar{G}_{ij}^{\;\;\; kl} \Box L_{kl}
   + 2\alpha\bar{R}^k_{\; (i} L_{j)k} - 2\alpha\bar{R}_{kilj} L^{kl} \nonumber\\
 &+& 2\nabbar_{(i} \alpha\nabbar^k L_{j)k} - 2\nabbar^k \left( \alpha_{(i} L_{j)k} \right)
   + 2\alpha^{k} \nabbar_{(i} L_{j)k} \nonumber\\
 &+& 2\left(\nabbar_{(i} \nabbar^k \alpha \right) L_{j)k}
  - \alpha\nabbar_i\nabbar_j L - \gbar_{ij} \nabbar^k\nabbar^l (\alpha L_{kl} ) \nonumber\\
 &-& \frac{1}{\alpha} \bar{G}_{ij}^{\;\;\; kl} \nabbar_{(k} \alpha^2\nabbar_{l)}
     \left( \frac{A}{\alpha} \right),
\label{Eq-WaveSij}
\eea
where the d'Alembertian is defined according to
\bdm
\Box = \frac{1}{\alpha}\p_t^{\; 2} - \nabbar^r\alpha\nabbar_r\, ,
\edm
and where
\bea
S^{BG}_{ij} &\equiv& \alpha G_{ij} L + \gbar_{ij} \alpha G^{kl} L_{kl} - \alpha R L_{ij}
 + \frac{1}{2}\gbar_{ij} \alpha R L \nonumber\\
&+& G_{ij} A.
\label{Eq-SBGDef}
\eea
In addition, we consider the equation
\bea
S_{00} &\equiv& \frac{-1}{\sqrt{-g}}\,\p_t\, \delta\left( \sqrt{-g}\, G_{00} \right) \label{Eq-WaveS00}\\
       &=& -\frac{1}{\alpha} G_{00} A 
  - \frac{1}{\alpha}\bar{G}^{ijkl} \nabbar_i \alpha^2 \nabbar_j \left( \frac{L_{kl}}{\alpha} \right) + G^{ij} L_{ij}\, .
\nonumber
\eea
As expected from the general formula (\ref{Eq-DeltaGmunuSym}), the
spatial part of the operators in $(S_{00}, S_{ij})$ is symmetric in
$u = (A, L_{ij})$ with respect to the inner product
\be
\sprod{u^{(1)}}{u^{(2)}} = \int\limits_{\Sigma} \left( A^{(1)} A^{(2)} 
 + \gbar^{ik}\gbar^{jl} L^{(1)}_{ij} L^{(2)}_{kl} \right) \etabar.
\label{Eq-ADMInnerProduct}
\ee
Furthermore, one has the following constraint equation for $L_{ij}$:
\be
\delta G_{i0} = \bar{G}_{ijkl} \nabbar^j \left( \frac{L^{kl}}{\alpha} \right).
\label{Eq-WaveConstraint}
\ee

The system $S_{ij}$ still has the following drawbacks:
First, the spatial operator is not elliptic: Considering, in particular, 
perturbations of the form 
$L_{kl} = \frac{1}{3}L \gbar_{ij}$, the second 
derivatives yield 
$-\frac{\alpha}{3}(\nabbar_i\nabbar_j - \gbar_{ij}\lapbar)L$,
which is not governed by an elliptic operator.
Second, the ``kinematic energy metric'', $\bar{G}^{ijkl}$, is not positive.
Indeed, for a symmetric tensor field $t_{ij} = \hat{t}_{ij} + \frac{1}{3} \gbar_{ij} t$,
where $\hat{t}_{ij}$ is trace-less, one finds
$\bar{G}_{ij}^{\;\;\; kl} t_{kl} = \hat{t}_{ij} - \frac{2}{3} \gbar_{ij} t$.
Finally, there exists no evolution equation for the perturbed lapse $A$.

In order to overcome these problems, we first apply ideas 
introduced in \cite{CBYA-Hyperbolic}, where it was shown that the spatial 
operator becomes elliptic after using the momentum constraint (\ref{Eq-WaveConstraint}) 
and its spatial derivatives. Slightly changing
the procedure outlined in \cite{CBYA-Hyperbolic}, this can be achieved
without destroying the symmetry of the final operator.
It turns out that the combination
\bea
\Lambda_{ij} &\equiv& \frac{\alpha}{\sqrt{-g}}\,\p_t\, \delta\left( \sqrt{-g}\, G_{ij} \right)
  - \frac{2}{\alpha} \nabbar_{(i} \left( \alpha^3 \delta G_{j)0} \right) \nonumber\\
 &+& \frac{1}{\alpha^2} \gbar_{ij} \nabbar^k \left( \alpha^4 \delta G_{k0} \right)
\label{Eq-LambdaDef}
\eea
almost yields the desired result:
$\Lambda_{ij} = \Lambda^{V}_{ij} + S^{BG}_{ij}$, where
\bea
\Lambda^{V}_{ij} &=&
\bar{G}_{ij}^{\;\;\; kl} \Box L_{kl}
 + 2\alpha\bar{R}^k_{\; (i} L_{j)k} - 2\alpha\bar{R}_{kilj} L^{kl} \nonumber\\
&+& 4\nabbar_{(i} \left( \alpha^k L_{j)k} \right) - 4\alpha_{(i} \nabbar^k L_{j)k} 
 - 2\alpha \nabbar^k\! \left( \frac{\alpha_{(i}}{\alpha} \right) L_{j)k}
\nonumber\\
&-& \frac{2}{\alpha}\nabbar_i( \alpha\alpha_j) L
 - \frac{2}{\alpha} \gbar_{ij} \nabbar^k (\alpha\alpha^l) L_{kl}
 - 2\alpha \gbar_{ij} \nabbar^k\! \left( \frac{\alpha_k}{\alpha} \right) L \nonumber\\
&+& \frac{1}{\alpha}\bar{G}_{ij}^{\;\;\; kl} \left[ \nabbar_{(k}\alpha^2\nabbar_{l)} \left( L 
 - \frac{1}{\alpha}A \right) \right].
\label{Eq-LambdaijDes}
\eea
(The main difference between $\Lambda_{ij}$ and the
$\Omega_{ij}$ introduced in \cite{CBYA-Hyperbolic} lies in
the symmetrizing terms.)

In order to obtain a symmetric wave equation, one makes
use of the residual gauge freedom (\ref{Eq-ADMResGaugeTrans})
and adopts one of the following two gauges:
The first possibility is to choose the {\it harmonic gauge\/}
\bdm
A = \alpha L,
\edm
in which case $\Lambda^{V}_{ij}$ becomes a symmetric operator in $L_{ij}$.
Furthermore, this also solves the lapse-problem.
Unfortunately, the De Witt metric still appears in front of the
d'Alembertian $\Box$, and, as a consequence, the spatial operator
is not elliptic.
Alternatively, one can multiply $\Lambda^{V}_{ij}$ from the left by
the inverse of the De Witt metric,
\bdm
\bar{G}^{-1}_{ijkl} = \frac{1}{2}\left( \gbar_{ik} \gbar_{jl} + \gbar_{il} \gbar_{jk}
- \gbar_{ij} \gbar_{kl} \right),
\edm
which yields a hyperbolic wave-operator, which is, however, only
symmetric with respect to the inner product induced by the indefinite
de Witt metric $\bar{G}^{ijkl}$. This possibility has been considered
in \cite{BHS-Letter}.

In order to get a positive kinematic energy metric, it seems unavoidable
to choose the {\it maximal \/} (or constant mean curvature) slicing
condition,
\bdm
L = \gbar^{ij} L_{ij} = 0.
\edm
(This gauge may not always exist if the background manifold $\Sigma$ is closed.
In fact, Eq. (\ref{Eq-ADMResGaugeTrans}) shows that one has to solve an elliptic
equation for $f$ in order to get $L=0$.)
Adopting the maximal gauge, the trace-less part of $\Lambda_{ij}$
yields a manifestly hyperbolic, self-adjoint operator for the symmetric, trace-less
tensor $L_{ij}$. However, the perturbations of the lapse are still present in this gauge.
The amplitude $A$ obeys an elliptic equation which corresponds to the
trace of $\Lambda^{V}_{ij}$.
Hence, the system is of mixed hyperbolic-elliptic type.
However, as will be shown below, all terms involving $A$ disappear
when the operator in $\Lambda_{ij}$ is protected onto the constraint manifold defined by
(\ref{Eq-WaveConstraint}).

\subsection{The linearized Bianchi identities}

In the following the linearized version of the Bianchi identities (\ref{Eq-ADMBianchi})
will be needed. In particular, we want to show explicitely that the
constraint equation (\ref{Eq-WaveConstraint}) propagates.
The ADM metric (\ref{Eq-ADMMetric}) with $\beta = 0$ has the
following Christoffel symbols:
\bea
&& \chris{t}{tt} = \frac{\dot{\alpha}}{\alpha}, \;\;\;
   \chris{t}{tj} = \frac{\alpha_j}{\alpha}, \;\;\;
   \chris{t}{ij} = \frac{1}{\alpha} K_{ij}, \nonumber\\
&& \chris{k}{tt} = \alpha\alpha^k, \;\;\;
   \chris{k}{tj} = \alpha K^k_{\; j}, \;\;\;
   \chris{k}{ij} = \chrisbar{k}{ij} \, .
\eea
Using this in Eq. (\ref{Eq-ADMBianchi}) yields the
$3+1$ decomposition of the Bianchi identities:
\bea
0 &=& \nabla^\mu G_{\mu t} \nonumber\\
  &=& -\dot{G}_{00} - \alpha K G_{00} + \frac{1}{\alpha} \nabbar^k (\alpha^2 G_{0k}) - \alpha G^{kl} K_{kl},
\label{Eq-ADMBianchit}\\
0 &=& \alpha\nabla^\mu G_{\mu j} \nonumber\\
  &=& -\dot{G}_{0j} + G_{00}\alpha_j - G_{0j}\alpha K + \nabbar^k (\alpha G_{kj} ).
\label{Eq-ADMBianchij}
\eea
Linearizing Eq. (\ref{Eq-ADMBianchit}) on a static background yields
\be
\delta\dot{G}_{00} - \frac{1}{\alpha} \nabbar^k (\alpha^2 \delta G_{0k} ) 
= -\alpha G^{kl}\delta K_{kl} - \alpha G_{00} \delta K,
\label{Eq-BianchiFirstId}
\ee
which shows that no new constraint equations for $\delta K_{ij}$ are obtained when the
Hamilton constraint is differentiated with respect to $t$.

Using Eq. (\ref{Eq-BianchiFirstId}), the linearized version of Eq. (\ref{Eq-ADMBianchij}) becomes 
\bea
0 &=& \Box C_j - \alpha \bar{R}_{jk} C^k
   + 2\nabbar_j \left( \alpha^k C_k \right) - 2\alpha_j \nabbar^k C_k \label{Eq-BianchiSecondID}\\
  &-& \alpha\nabbar^k \left( \frac{\alpha_k}{\alpha} \right) C_j
   - \frac{1}{\alpha}\nabbar^k(\alpha\alpha_j) C_k
   - \alpha\nabbar^k \Lambda_{kj} + (BG), \nonumber
\eea
where $\Lambda_{ij}$ is given in (\ref{Eq-LambdaDef}).
The constraint variables $C_j$ are defined according to
\bdm
C_j \equiv \alpha^2 \delta G_{0j} \, ,
\edm
and $(BG)$ denotes terms which are proportional to components of
the Einstein tensor. The fact that
Eq. (\ref{Eq-BianchiSecondID}) also holds for the stress-energy tensor 
if the matter equations are satisfied, implies that these terms need not be
considered any longer, provided that $\delta G_{\mu\nu}$ is replaced by
$\delta G_{\mu\nu} - 8\pi G \delta T_{\mu\nu}$ in Eqs. (\ref{Eq-BianchiSecondID})
and the definition of the $C_j$.

Hence, if the evolution equations, $\Lambda_{ij}=0$, and the background equations
are fulfilled, it follows that the constraints $C_j$
propagate. Note that the operator governing the $C_j$ in Eq. (\ref{Eq-BianchiSecondID})
is hyperbolic and symmetric.

\subsection{Summary}

The linearized Einstein equations for a static background yield
a hyperbolic, formally self-adjoint operator for the variation of
the extrinsic curvature. This quantity is coordinate-invariant with
respect to infinitesimal diffeomorphisms within the slices orthogonal
to $\p_t$. The wave operator is constructed from the combination
$\Lambda_{ij}$ defined in (\ref{Eq-LambdaDef}).
In the maximal slicing gauge, $\delta K = 0$, this yields
(using also the background equation $R_{00} = \frac{1}{\alpha}\lapbar\alpha$)
\be
\Lambda^k_{\; k} = \gbar^{kl} S^{(BG)}_{kl} + 2\left(\lapbar - R_{00} \right) A
 - \frac{4}{\alpha} \nabbar^k (\alpha\alpha^l) L_{kl}\, ,
\label{Eq-WaveAlpha}
\ee
and
\bea
\hat{\Lambda}_{ij} &=& \hat{S}^{(BG)}_{ij} + \Box L_{ij}
    + 2\alpha\bar{R}^k_{\; (i} L_{j)k} - 2\alpha\bar{R}_{kilj} L^{kl} \nonumber\\
   &+& 4\nabbar_{(i} \left( \alpha^k L_{j)k} \right) - 4\alpha_{(i} \nabbar^k L_{j)k} 
    + 2\alpha\nabbar^k\! \left( \frac{\alpha_{(i}}{\alpha} \right) L_{j)k} \nonumber\\
   &-& \frac{1}{\alpha} \nabbar_{(i}\alpha^2\nabbar_{j)} \left( \frac{A}{\alpha} \right) \nonumber\\
   &+& \frac{1}{3} \gbar_{ij} \left( -\frac{2}{\alpha} \nabbar^k(\alpha\alpha^l) L_{kl} + \lapbar A
    - R_{00} A \right),
\label{Eq-WaveLij}
\eea
where $\hat{\Lambda}_{ij}$ denotes the trace-less part of $\Lambda_{ij}$,
and where the trace-less symmetric tensor $L_{ij}$ and
the scalar $A$ are defined by
\bdm
L_{ij} \equiv \alpha\delta K_{ij}, \;\;\;
A \equiv \delta\dot{\alpha}.
\edm
These quantities are subject to the residual gauge transformation
(\ref{Eq-ADMResGaugeTrans}) with $\nabbar^k (\alpha^2\nabbar_k f) = 0$.
The system $(\hat{\Lambda}_{ij}, \Lambda^k_{\; k})$ provides
a closed set of evolution equations for $(L_{ij}, A)$, which
is manifestly hyperbolic and symmetric in $L_{ij}$ and elliptic
in $A$. The operator $-\lapbar + R_{00}$ in equation
(\ref{Eq-WaveAlpha}) is non-negative if the strong energy
condition and the background equations hold, i.e., if
\mbox{$R_{00} \geq 0$}.

Furthermore, $L_{ij}$ obeys the momentum constraint equation
\bdm
\delta G_{i0} = \nabbar^j \left( \frac{L_{ij}}{\alpha} \right).
\edm
The constraint variables,
\bdm
C_j = \alpha^2 \left( \delta G_{0j} - 8\pi G T_{0j} \right),
\edm
propagate by virtue of Eq. (\ref{Eq-BianchiSecondID}).

Finally, the differentiated Hamilton constraint yields
\bea
\Lambda_{00} &\equiv& \frac{-1}{\sqrt{-g}}\, \p_t\, \delta\left( \sqrt{-g}\, G_{00} \right) \nonumber\\
             &=& -\frac{1}{\alpha} G_{00} A - \frac{1}{\alpha}\nabbar^i \alpha^2 \nabbar^j \left( \frac{L_{ij}}{\alpha} \right) + G^{ij} L_{ij},
\label{Eq-Lambda00Def}
\eea
and the system $(\Lambda_{00}, \hat{\Lambda}_{ij})$ is
symmetric in $(A, L_{ij})$ with respect to the inner product
defined in (\ref{Eq-ADMInnerProduct}).

\section{The vacuum equations}
\label{Sect-ADMVacEq}

We start by analyzing the wave operator for
vacuum perturbations of a static vacuum space-time.
By virtue of the constraint equation
\be
0 = \nabbar^j \left( \frac{L_{ij}}{\alpha} \right),
\label{Eq-WaveConsVac}
\ee
the equation $\Lambda_{00} = 0$ is fulfilled and 
Eq. (\ref{Eq-WaveAlpha}) can be rewritten as
\be
\lapbar A = 2\nabbar^k (\alpha^l L_{kl}).
\label{Eq-WaveAlphaVac}
\ee
The Laplacian on the left-hand side (LHS) is elliptic and
symmetric on the dense subspace of 
$C^{\infty}$-functions in the Hilbert space $L^2(\Sigma, \etabar)$.
As a consequence, its image is equal to the orthogonal complement of its kernel.
If $\Sigma$ is compact, the kernel $\ker\lapbar$ is the set
of constant functions on $\Sigma$. If $\Sigma$ is not compact,
but all perturbations vanish sufficiently fast at space-like
infinity, $\ker\lapbar$ is trivial.
In either case, the RHS of (\ref{Eq-WaveAlphaVac}) is orthogonal
to $\ker\lapbar$ and therefore lies in the image of $\lapbar$.
This shows that equation (\ref{Eq-WaveAlphaVac}) is solvable.
A solution, which is unique up to the addition of an element in
$\ker\lapbar$, can be written formally as
\be
A = 2\ul{\lapbar}^{-1} \nabbar^k (\alpha^l L_{kl}),
\label{Eq-AVac}
\ee
where $\ul{\lapbar}$ denotes the restriction of $\lapbar$ on
the orthogonal complement of $\ker\lapbar$.

\subsection{The initial value formulation}
\label{Sect-ADMVacInit}

A necessary condition for Einstein's equations to hold is
that the evolution equations $\hat{\Lambda}_{ij} = 0$
and the constraint equations (\ref{Eq-WaveConsVac}) are satisfied.
In order to find a solution of Einstein's equations, these equations must, 
however, be supplemented with
\bea
0 &=& \delta G_{00} = \frac{1}{2} \delta\bar{R} =
\frac{1}{2\alpha} \bar{G}^{ijkl} \nabbar_i\alpha^2\nabbar_j \left( \frac{\delta\gbar_{kl}}{\alpha} \right),
\label{Eq-WaveG00}\\
0 &=& \delta G_{ij} = \delta\bar{G}_{ij}
   - \delta\,\frac{1}{\alpha}\bar{G}_{ij}^{\;\;\; kl} \nabbar_k\nabbar_l\alpha
   + \frac{1}{\alpha^2} \dot{L}_{ij}\, .
\label{Eq-WaveGij}
\eea
The first equation is the linearized Hamilton constraint, while the
second equation is the evolution equation which earlier was differentiated
with respect to $t$ to construct the wave operator.

The linearized Einstein equations are solved as follows:
\begin{enumerate}
\item{Specify any $3$-metric $\delta\gbar_{ij} = \delta\gbar(t=0)_{ij}$
and any symmetric, trace-less tensor field $L_{ij}$ on an initial
time-slice $\Sigma_0$, such that the Hamilton-constraint
(\ref{Eq-WaveG00}) and the momentum constraint (\ref{Eq-WaveConsVac}) hold.}
\item{Choose any convenient function $\delta\alpha = \delta\alpha(t=0)$ on $\Sigma_0$.}
\item{Compute $A$ and $\dot{L}_{ij}$ on $\Sigma_0$ from equation
(\ref{Eq-AVac}) and (\ref{Eq-WaveGij}), respectively.
By virtue of the background equations and the linearization of the
Bianchi identity (\ref{Eq-ADMBianchij}), 
$\dot{L}_{ij}/\alpha$ is trace- and divergence-free. Therefore, the
momentum constraint propagates as a consequence of Eq. (\ref{Eq-BianchiSecondID}).}
\item{Evolve $(A, L_{ij})$ via the equation $\hat{\Lambda}_{ij} = 0$
and equation (\ref{Eq-AVac}).}
\item{The linearized metric coefficients are eventually obtained from
\bea
&& \delta\alpha(t) = \delta\alpha(0) + \int\limits_0^t A(\tau) \diff\tau, \nonumber\\
&& \delta\gbar(t)_{ij} = \delta\gbar(0)_{ij} + \int\limits_0^t 2L(\tau)_{ij} \diff\tau. \nonumber
\eea
}
\end{enumerate}
Note that in contrast to the full non-linear constraint equations,
the linearized Hamilton and momentum constraint decouple. (This
follows from the fact that $\Ktens$ vanishes on the background.)
The ``linearized'' conformal method to solve the Hamilton
constraint is to split
\bdm
\delta\gbar_{ij} = \hat{m}_{ij} + \frac{1}{3}\gbar_{ij} \phi
\edm
in its trace and trace-less parts.
Inserting this into Eq. (\ref{Eq-WaveG00}) yields
\bdm
\frac{1}{\alpha}\nabbar^k \alpha^2 \nabbar_k \left( \frac{\phi}{\alpha} \right)
= \frac{3}{2}\,\frac{1}{\alpha} \nabbar^k \alpha^2 \nabbar^l \left( \frac{\hat{m}_{kl}}{\alpha} \right).
\edm
Again, the RHS of this equation is orthogonal to the kernel
of the operator on the LHS, which is symmetric and elliptic.
Therefore, the trace-less part $\hat{m}_{ij}$ can be
specified freely, and the above equation can be solved for $\phi$.
How to solve the linearized momentum constraint is explained below.

\subsection{The projection onto the constraint manifold}

Our next aim is to extract the ``pure dynamical degrees of freedom''
by solving the momentum constraint, and to project the wave operator
onto the space of dynamical variables. For vacuum perturbations this can be
achieved by using the ``York decomposition'' \cite{Y-Decomp}.

First, we replace $L_{ij}$ by
\bdm
Z_{ij} = \frac{1}{\sqrt{\alpha}} L_{ij}\, ,
\edm
and the operator $\hat{\Lambda}_{ij}$ by $\sqrt{\alpha}\hat{\Lambda}_{ij}$.
This eliminates the factor $1/\alpha$ in front of the second
time derivatives in $\hat{\Lambda}_{ij}$. The linearized momentum
constraint now reads
\bdm
\nabbar^j \left( \frac{Z_{ij}}{\sqrt{\alpha}} \right) = 0.
\edm
Let ${\cal T}$ denote the space of all $C^\infty$ symmetric, trace-less
covariant tensor fields on $\Sigma$, and let ${\cal V}$ denote
the space of all $C^\infty$ vector fields on $\Sigma$. We assume
that either $\Sigma$ is compact or all tensor fields vanish
sufficiently rapid on the ``border'' of $\Sigma$.
Consider the operator
\bea
&& \bbL^\dagger : {\cal T} \longrightarrow {\cal V}, \nonumber\\
&& (\bbL^\dagger Z)_i = 2\nabbar^j\left( \frac{Z_{ij}}{\sqrt{\alpha}} \right), \nonumber
\eea
and let $W\in {\cal V}$ and $Z\in {\cal T}$. With respect to the
inner products on ${\cal V}$ and ${\cal T}$ induced by $\gbar_{ij}\,$,
we have
\bdm
\sprod{W}{\bbL^\dagger Z} = \sprod{\bbL W}{Z},
\edm
where $\bbL : {\cal V} \longrightarrow {\cal T}$ is given by
\bdm
(\bbL W)_{ij} = -\frac{1}{\sqrt{\alpha}}
\left( \nabbar_i W_j + \nabbar_j W_i - \frac{2}{3}\gbar_{ij} \nabbar^k W_k \right).
\edm
The operator $\bbL^\dagger \bbL : {\cal V} \longrightarrow {\cal V}$
is symmetric, positive semi-definite and elliptic.
Its kernel $\ker( \bbL^\dagger \bbL ) = \ker\bbL$ consists of
all conformal Killing vector fields on $\Sigma$.
As a consequence, we can solve the equation
\bdm
\bbL^\dagger \bbL W = \bbL^\dagger Z,
\edm
for given $Z\in {\cal T}$,
\bdm
W = \ul{ (\bbL^\dagger \bbL) }^{-1} \bbL^\dagger Z + W_0,
\edm
where $\ul{ (\bbL^\dagger \bbL) }$ denotes the restriction
of $\bbL^\dagger \bbL$ on the orthogonal complement of its
kernel, and where $W_0\in\ker\bbL$.

The orthogonal projector on the constraint manifold
\bdm
{\cal C} 
 = \left\{ (Z_{ij})\in {\cal T} \Bigm| \nabbar^j \left( \frac{Z_{ij}}{\sqrt{\alpha}} \right) = 0 \right\}
 = \ker\bbL^\dagger
\edm
can therefore be represented as
\bdm
\bbP = \identy - \bbL \ul{ (\bbL^\dagger \bbL) }^{-1} \bbL^\dagger.
\edm
Any symmetric, trace-less tensor field $Z\in {\cal T}$ can
be decomposed according to
\bdm
Z_{ij} = Z^{TT}_{ij} + Z^{TL}_{ij},
\edm
where $Z^{TT}/\sqrt{\alpha} = \bbP Z/\sqrt{\alpha}$ is trace-less and
divergence-free, and where
\bdm
Z^{TL}_{ij} = \bbL \ul{ (\bbL^\dagger \bbL) }^{-1} \bbL^\dagger Z_{ij} \equiv (\bbL W)_{ij}
\edm
is the trace-less longitudinal part.
This decomposition is covariant and orthogonal with respect
to the metric on ${\cal T}$ induced by $\gbar_{ij}$.
For $\alpha=1$, it corresponds to York's decomposition
for symmetric tensor fields $K_{ij}$. York's decomposition
has further conformal properties which play an essential
role in solving the full non-linear constraint equations
\cite{Y-Decomp}.

In our case, the linearized momentum constraint is
solved by taking
\bdm
Z_{ij} = \bbP \tilde{Z}_{ij},
\edm
where $\tilde{Z}_{ij}$ is any symmetric, trace-less
tensor field on $\Sigma$. Of course, this requires the
computation of $\ul{ (\bbL^\dagger \bbL) }^{-1}$, which
is a non-local operator.

Consider now the wave operator $\sqrt{\alpha}\hat{\Lambda}_{ij}$ defined
in (\ref{Eq-WaveLij}). It has the form
\bdm
\sqrt{\alpha}\hat{\Lambda}_{ij} = \left( \p_t^{\, 2} + {\cal A} \right) Z_{ij} + b_{ij}(A),
\edm
where ${\cal A}$ is a spatial, elliptic and self-adjoint operator and
the functions $b_{ij}(A)$ are given by
\be
b_{ij}(A) = -\frac{1}{\sqrt{\alpha}} \nabbar_{(i}\alpha^2\nabbar_{j)} \left( \frac{A}{\alpha} \right)
            + \frac{\sqrt{\alpha}}{3} \gbar_{ij} \lapbar A.
\ee
By virtue of the linearized Bianchi identity (\ref{Eq-BianchiSecondID}),
it follows that the spatial operator defined by
\bdm
Z_{ij} \mapsto {\cal A} Z_{ij} + b_{ij}(A),
\edm
where $A$ is given by (\ref{Eq-AVac}),
maps the constraint manifold ${\cal C}$ into itself.
(In particular, this implies that the momentum constraint propagates.)
As a consequence, we have
\bdm
{\cal A} \bbP Z_{ij} + b_{ij}(A) = \bbP\left( {\cal A} \bbP Z_{ij} + b_{ij}(A) \right)
\edm
(still provided that equation (\ref{Eq-AVac}) holds).
Now it easy to see that the $b_{ij}$ map into a space
which is orthogonal to the constraint manifold.
Indeed, the spatial operator in the system
$\sqrt{\alpha}(\Lambda_{00},\hat{\Lambda}_{ij})$ can be shown to have the form
\bdm
\left( \begin{array}{cc} 0 & b^\dagger \\ b & {\cal A} \end{array} \right).
\edm
As mentioned above, $b^\dagger\bbP = 0$, also implying that
$\bbP b = 0$. Thus, the wave equation restricted to the constraint
manifold takes the form of a symmetric wave equation,
\bdm
\left( \bbP\p_t^{\, 2} + \tilde{{\cal A}} \right) Z_{ij} = 0,
\edm
where $\tilde{{\cal A}} \equiv \bbP {\cal A} \bbP$ is
a (formally) self-adjoint spatial operator.

In conclusion, we have shown that the wave operator, projected
onto the (momentum) constraint manifold, yields a
symmetric wave equation for the symmetric, trace-less
tensor $Z_{ij} = \sqrt{\alpha} \delta K_{ij}$. As expected,
the variation of the lapse does not appear in this equation.
The following two difficulties remain:
First, the quantity $Z_{ij}$ is not coordinate-invariant
with respect to reparametrization of time. Hence, we have
not yet isolated all physical degrees of freedom.
The second difficulty is to find an explicit characterization
of the constraint manifold in order to have an explicit representation
of the operator $\tilde{{\cal A}}$.
We have not yet solved these problems for an arbitrary static
background. However, the additional structure provided by
a spherically symmetric background, enables one to solve both problems.
This is discussed in Appendix \ref{App-A}, where a natural
derivation of the Regge-Wheeler and Zerilli equations is given.

\subsection{Locally flat space-times}

To conclude this section, we briefly specialize to the case
where $\alpha \equiv 1$. Using the background equations
(\ref{Eq-ADMBack}), this implies that $(\Sigma, \gbartens)$
is locally flat.
Equation (\ref{Eq-WaveAlpha}) then yields $A = A(t)$,
and the pulsation equation reduces to
\bdm
0 = \hat{\Lambda}_{ij} = \left( \p_t^{\; 2} - \lapbar \right) L_{ij}\, ,
\edm
with the constraint
\bdm
0 = \nabbar^j L_{ij}\, .
\edm
These equations are well-known from the weak-field limit
of general relativity. They can be solved by Fourier transformation.
In particular, it follows that all vacuum space-times, which can
be represented by $I\times \Sigma$ with $I\subseteq\RR$ and
$\Sigma$ locally flat, are linearly stable.

\section{The coupling to Yang-Mills-Higgs fields}

An important new feature of the formulation presented in this 
paper is the fact that it extends to gravitating matter fields
in a natural way. As mentioned earlier, this is due to 
that - in contrast to the traditional metric approach - the wave 
operator governing the variation of the extrinsic curvature 
appears already off-shell.

In this section, we show that the equations governing fluctuations
of a static solution to the Einstein-Yang-Mills-Higgs (EYMH) equations
assume the form of a symmetric wave equation. While this is explicitely
established for the triplet case, we emphasize that it holds true
whenever the gauge group is a compact Lie group.
(Compactness guarantees the existence of an Ad-invariant
scalar product on the Lie algebra.)

\subsection{The ADM equations}

In the ADM formalism, the gauge potential $A$ is parametrized
in terms of a scalar field $\Phi \equiv -i_{\p_n} A$ and a one-form $\Abar$
(both Lie algebra valued),
\bdm
A = -\Phi\, \alpha\diff t + \Abar_i \left( \diff x^i + \beta^i \diff t \right).
\edm
Similarly to the extrinsic curvature, $K_{ij} = \frac{1}{2} L_n g_{ij}$,
in the gravitational case, the electric one-form is defined by
\bdm
E \equiv -i_n F,
\edm
where $F = \diff A + A \wedge A$ is the YM field strength.
We also define the momentum belonging to the Higgs field
$H$ by
\bdm
\Pi \equiv i_n \Diff H,
\edm
where $\Diff\equiv \diff + [A |\;\; ]$ denotes the covariant
derivative with respect to $A$.

In terms of these quantities, the ADM decomposition of the YMH
equations yields the following equations (which we list here
only for vanishing shift, $\beta = 0$, since they will be needed
only for a static background and in a
gauge with $\delta\beta=0$; the general case is obtained after
the substitution $\p_t\mapsto \p_t - \bar{L}_\beta$.):
The Gauss constraint is
\bdm
-(\ast\Diff\ast F - J)_0 = \ast\Diffbar\ast E - [ H,\Pi].
\edm
The evolution equations comprise the definition
of $E$ and $\Pi$,
\bea
E   &=& -\frac{1}{\alpha} \left( \p_t \Abar + \Diffbar(\alpha \Phi) \right), \nonumber\\
\Pi &=&  \frac{1}{\alpha} \p_t  H - [\Phi,  H], \nonumber
\eea
respectively, and the equations
\bea
 -\alpha(\ast\Diff\ast F - J)^i &=& \frac{1}{\sqrt{\gbar}}\,\p_t \left( \sqrt{\gbar}\, E^i \right)
 - \astbar\Diffbar\astbar\left( \alpha\bar{F} \right)^i \nonumber\\
&-& [\alpha \Phi, E^i ] - [\alpha H, \Diffbar^i H], \nonumber\\
 \alpha\left( \ast\Diff\ast\Diff H + 2V' H \right) &=& \frac{1}{\sqrt{\gbar}}\,\p_t \left( \sqrt{\gbar}\,\Pi \right)
 - \astbar\Diffbar\astbar\left( \alpha\Diffbar H \right) \nonumber\\
&-& [\alpha \Phi, \Pi ] + 2\alpha V' H. \nonumber
\eea
All quantities with a bar refer to the $3$-metric $\gbartens$
and to the magnetic part of the potential, $\Abar$.
Also, $E^i \equiv \gbar^{ij} E_j$ is raised with the
$3$-metric.
The Higgs potential $V$ is assumed to be a function of
$|H|^2 = \Trace(H^2)$ only, with derivative $V'$.

There is no evolution equation for the electric potential $\Phi$,
which reflects the invariance of the theory with respect to
gauge transformations of the gauge potential $A$.
In fact, $\Phi$ plays a similar role as the shift $\beta$
for the gravitational ADM equations. 
It is easy to see that the Gauss constraint propagates
as a consequence of the identity $\Diff^2\ast F = [F | \ast F] = 0$
and the corresponding consistency condition
$\Diff\ast J = -[\Diff H | \ast\Diff H] - [ H,\Diff\ast\Diff H] = 0$
for the matter current, which holds by virtue of the Higgs equations.

In order to solve the coupled EYMH equations, 
the 3+1 decomposition of the stress-energy tensor
$T_{\mu\nu} = T^{(YM)}_{\mu\nu} + T^{(H)}_{\mu\nu}$ is needed:
\bea
T^{(YM)}_{00} &=& \frac{1}{4\pi}\Trace\Big\{ \frac{1}{2} E_k E^k + \frac{1}{4}\bar{F}_{kl}\bar{F}^{kl} \Big\}, \nonumber\\
T^{(YM)}_{i0} &=& \frac{1}{4\pi}\Trace\Big\{ E^k \bar{F}_{ki} \Big\}, \nonumber\\
T^{(YM)}_{ij} &=& \frac{1}{4\pi}\Trace\Big\{ -E_i E_j + \frac{1}{2}\gbar_{ij} E_k E^k \nonumber\\
&& + \bar{F}_{ik}\bar{F}_j^{\; k} - \frac{1}{4}\gbar_{ij}\bar{F}_{kl}\bar{F}^{kl} \Big\}, \nonumber\\
T^{(H)}_{00} &=& \frac{1}{4\pi}\Trace\Big\{ \frac{1}{2}\Pi^2 
  + \frac{1}{2}(\Diffbar_k H)(\Diffbar^k H) \Big\} + \frac{1}{4\pi} V, \nonumber\\
T^{(H)}_{i0} &=& \frac{1}{4\pi}\Trace\Big\{ \Pi\,(\Diffbar_i H) \Big\}, \nonumber\\
T^{(H)}_{ij} &=& \frac{1}{4\pi}\Trace\Big\{ (\Diffbar_i H)(\Diffbar_j H)
  - \frac{1}{2}\gbar_{ij}(\Diffbar_k H)(\Diffbar^k H) \nonumber\\
&& + \frac{1}{2}\gbar_{ij}\Pi^2 \Bigr\} - \frac{1}{4\pi}\gbar_{ij} V. \nonumber
\eea
Here, $\Trace$ denotes an Ad-invariant innerproduct on
the Lie algebra. The complete set of EYMH equations in the ADM formalism
is now obtained from the expressions
(\ref{Eq-ADMConsHam}), (\ref{Eq-ADMConsMom}) and (\ref{Eq-ADMEvol})
for the Einstein tensor, and the definition (\ref{Eq-ADMExtrinsic})
of the extrinsic curvature.

\subsection{Gauge-invariant quantities}

For a static background the slicing may be chosen such that
$\gtens = -\alpha^2\diff t^2 + \gbartens$ and $K_{ij} = 0$.
As a consequence, $G_{i0} = 0$, and it is consistent to
set $E = 0$ and $\Pi = 0$. Hence, we are considering a static, purely
magnetic background:
\bdm
A = \Abar_i\diff x^i ,
\edm
where $\Abar$ and $ H$ do not depend on $t$.
The YMH equations reduce to
\bea
\astbar\Diffbar\astbar\left( \alpha\bar{F} \right) + [\alpha H, \Diffbar^i H] &=& 0,
\nonumber\\
\astbar\Diffbar\astbar\left( \alpha\Diffbar H \right) - 2\alpha V' H &=& 0.
\label{Eq-ADMBackYMH}
\eea

The perturbed amplitudes, $\delta \Phi$, $\delta E$,
$\delta H$ and $\delta\Pi$ are subject to both coordinate
and gauge transformations. Since $E$ and $\Pi$ vanish
on the background, we except them to be gauge-invariant to
first order:
Under infinitesimal coordinate transformations generated by
a vector field $X = (f, X^i)$, one finds, using the same
formulae as in section \ref{Sect-ADMCoordInv}:
\bea
\delta E_i &\mapsto& \delta E_i + \alpha\bar{F}_{ij} f^j, \nonumber\\
\delta \Pi &\mapsto& \delta\Pi + \alpha(\Diffbar_j H) f^j \nonumber.
\eea
Therefore, $\delta E$ and $\delta\Pi$ are vector-invariant
quantities.
Under an infinitesimal gauge transformation generated by
a Lie-algebra valued scalar $\chi$, we find that
\bdm
\delta \Phi \mapsto \delta \Phi - \frac{1}{\alpha}\dot{\chi},
\edm
whereas $\delta E$ and $\delta\Pi$ remain invariant.
Hence, the gauge-invariant quantities to
evolve are the perturbed {\it electric one-form\/} and
the perturbed {\it Higgs momentum\/}.

\subsection{The linearized equations}

In order to simplify the derivation of the wave operator,
it is convenient to choose the gauge function $\chi$ such that
$\delta \Phi = 0$. In this gauge, the linearized ADM Gauss constraint becomes
\bdm
-\delta(\ast\Diff\ast F - J)_0 = \astbar\Diffbar\astbar\delta E - [ H, \delta\Pi],
\edm
while the ADM evolution equations are
\bdm
\delta E = -\frac{1}{\alpha} \p_t\, \delta\Abar, \;\;\;
\delta\Pi =  \frac{1}{\alpha} \p_t\, \delta H,
\edm
and
\bdm
-\alpha\delta(\ast\Diff\ast F - J)^i = \delta\dot{E}^i - \delta\astbar\Diffbar\astbar\left( \alpha\bar{F} \right)^i
-\delta[\alpha H, \Diffbar^i H],
\edm
\bea
\alpha\delta\left( \ast\Diff\ast\Diff H + 2V' H \right) &=& \delta\dot{\Pi}
 - \delta\astbar\Diffbar\astbar\left( \alpha\Diffbar H \right) \label{Eq-YMHADMEvol}\\
&+& 2\delta\left( \alpha V' H \right), \nonumber
\eea
These equations have exactly the same structure as the
gravitational equations (\ref{Eq-DeltaADMCons}),
(\ref{Eq-DeltaADM1}) and (\ref{Eq-DeltaADM2}), where
$\delta\dot{\Abar}_i$, $\delta\dot{ H}$, $\delta\dot{\gbar}_{ij}$,
as well as $\delta E_i$, $\delta\Pi$, $\delta K_{ij}$ correspond
to each other.
In order to obtain a wave equation for $\delta E$ and $\delta\Pi$,
we therefore differentiate the two equations in (\ref{Eq-YMHADMEvol})
with respect to $t$. Again defining vector-invariant quantities
according to
\bdm
{\cal E} \equiv \alpha\delta E = -\p_t\,\delta\Abar, \;\;\;
\Psi \equiv -\alpha\delta\Pi = -\p_t\, \delta H,
\edm
and also using 
\bdm
\p_t\,\delta\gbar_{ij} = 2L_{ij}, \;\;\;
\p_t\,\delta\alpha = A,
\edm
yields the following expressions:
\bea
&& -\alpha\p_t \delta(\ast\Diff\ast F - J)_i \nonumber\\
&=& \frac{1}{\alpha}\ddot{{\cal E}}_i + \astbar\Diffbar\astbar \alpha\Diffbar {\cal E}_i
 -\alpha [ \bar{F}_{ij}, {\cal E}^j] + \alpha\bigl[  H, [{\cal E}_i,  H] \bigr] \nonumber\\
&+& \alpha\Diffbar_i\biggl( [ H,\Psi] \biggr) - 2\alpha[ \Diffbar_i H, \Psi] \label{Eq-ADMLYMH1}\\
&-& 2\Diffbar^j\left( L_{kj} \alpha\bar{F}^k_{\; i} \right) + 2\alpha\bar{F}^{kj}\nabbar_j L_{ki}
 -\alpha\bar{F}_{ij} \nabbar^j \left( L + \frac{A}{\alpha} \right), \nonumber
\eea
and
\bea
&& -\alpha\p_t \delta\left( \ast\Diff\ast\Diff H + 2V' H \right) \nonumber\\
&=& \frac{1}{\alpha}\ddot{\Psi} - \astbar\Diffbar\astbar\alpha\Diffbar\Psi
 + 4\alpha V'' \Trace( H\Psi) H + 2\alpha V' \Psi \nonumber\\
&-& [ \Diffbar_k( \alpha {\cal E}^k),  H] - 2\alpha[ {\cal E}^k, \Diffbar_k H] \label{Eq-ADMLYMH2}\\
&-& 2\Diffbar^i\left( L_{ij} \alpha\Diffbar^j H \right)
 +\alpha(\Diffbar_j H)\left( L + \frac{A}{\alpha} \right), \nonumber
\eea
for the linearized YM equations and the linearized
Higgs equations, respectively.
Here, we have also used the background equations
(\ref{Eq-ADMBackYMH}). Equations (\ref{Eq-ADMLYMH1})
and (\ref{Eq-ADMLYMH2}) provide a system of evolution
equations for ${\cal E}$ and $\Psi$, which is manifestly
{\it symmetric\/} with respect to the inner product
\bea
&& \sprod{ ({\cal E}^{(1)}, \Psi^{(1)}) }{ ({\cal E}^{(2)}, \Psi^{(2)})} \nonumber\\
&& \equiv \int\limits_\Sigma \Trace\left\{ \gbar^{ij}{\cal E}^{(1)}_i {\cal E}^{(2)}_j 
 + \Psi^{(1)} \Psi^{(2)} \right\} \etabar.
\label{Eq-ADMYMInnerProduct}
\eea
However, we are again faced with the problem that
the evolution equation governing ${\cal E}$ is
{\it not hyperbolic\/}. Like in the gravitational case,
a hyperbolic wave operator can be constructed
without loosing the symmetry. Again, the strategy is to add
suitable combinations of the linearized Gauss constraint,
\be
-\alpha\delta(\ast\Diff\ast F - J)_0 =
\alpha\astbar\Diffbar\astbar\left( \frac{ {\cal E} }{\alpha} \right) + [ H, \Psi],
\label{Eq-LADMEYMHGaussCons}
\ee
and its spatial derivatives.
Furthermore, one can use the linearized momentum constraint,
\bea
C_i &\equiv& \alpha^2\delta( G_{i0} - 8\pi G T_{i0} ) \label{Eq-LADMEYMHMomentumCons}\\
    &=& \alpha^2\bar{G}_{ijkl} \nabbar^j \left( \frac{L^{kl}}{\alpha} \right)
     - 2G\alpha\Trace\left( \bar{F}^k_{\; i}{\cal E}_k - (\Diffbar_i H)\Psi \right),
\nonumber
\eea
in order to eliminate the divergence terms of $L_{ij}$ in the last lines of
Eqs. (\ref{Eq-ADMLYMH1}) and (\ref{Eq-ADMLYMH2}).
The motivation for this (besides eliminating first
order derivatives) is the following:
We will see that the adjoint of the operator governing
the matter perturbation in the linearized stress-energy tensor
contains no divergence term. In order to formulate the linearized
EYMH equations as a symmetric wave equation, no divergences
of $L_{ij}$ must therefore appear in the YMH equations.
The correct combinations turn out to be
\bea
\Lambda^{(YM)}_i &\equiv& -\alpha\p_t \delta(\ast\Diff\ast F - J)_i \nonumber\\
                 &+& \frac{1}{\alpha}\Diffbar \alpha^3\delta(\ast\Diff\ast F - J)_0
                   +2\bar{F}^k_{\;\, i} C_k, \nonumber\\
\Lambda^{(H)} &\equiv& -\alpha\p_t \delta\left( \ast\Diff\ast\Diff H + 2V' H \right) \nonumber\\
              &-& \alpha^2 \bigl[ \delta(\ast\Diff\ast F - J)_0,  H \bigr]
                 +2(\Diffbar^k H) C_k.
\label{Eq-LambdaYMHDef}
\eea
This yields the following hyperbolic, symmetric equations:
\bea
\Lambda^{(YM)}_i &=& \Box {\cal E}_i -\alpha [ \bar{F}_{ij}, {\cal E}^j] + \alpha\bigl[  H, [{\cal E}_i,  H] \bigr]
\nonumber\\
   &+& 4G\alpha\Trace\left( \bar{F}^l_{\; k}{\cal E}_l - \Diffbar_k H \Psi \right)\bar{F}_i^{\; k}
    - 2[ \Diffbar_i( \alpha H), \Psi] \nonumber\\
   &+& 2\alpha\bar{F}^{jk}\nabbar_k L_{ij} -\frac{2}{\alpha} L_{kj} \Diffbar^j(\alpha^2\bar{F}^{ki}) \nonumber\\
   &+& \alpha\bar{F}_{ij}\nabbar^j\left( L - \frac{A}{\alpha} \right) + 2\alpha^k\bar{F}_{ki} L,
\label{Eq-LEYMHADMMatter1}
\eea
\bea
\Lambda^{(H)} &=& \Box \Psi + 4\alpha V''\Trace( H\Psi) H + 2\alpha V'\Psi + \alpha\bigl[ [ H,\Psi],  H \bigr] \nonumber\\
   &-& 4G\alpha\Trace\left( \bar{F}^l_{\; k}{\cal E}_l - \Diffbar_k H \Psi \right)\Diffbar^k H
    - 2[ {\cal E}^k, \Diffbar_k(\alpha H) ] \nonumber\\
   &-& \frac{2}{\alpha} L_{jk} \Diffbar^j\left( \alpha^2\Diffbar^k H \right) \nonumber\\
   &-& \alpha(\Diffbar_j H)\nabbar^j\left( L - \frac{A}{\alpha} \right) +  2\alpha^k(\Diffbar_k H) L,
\label{Eq-LEYMHADMMatter2}
\eea
where the d'Alembertians are defined by
\bea
\Box {\cal E}_i &=& \frac{1}{\alpha}\p_t^{\, 2} {\cal E}_i + \Diffbar^\dagger\alpha\Diffbar {\cal E}_i
    + \frac{1}{\alpha}\Diffbar\alpha^3\Diffbar^\dagger\left( \frac{ {\cal E}_i}{\alpha} \right), \nonumber\\
\Box \Psi &=& \frac{1}{\alpha}\p_t^{\, 2}\Psi + \Diffbar^\dagger\alpha\Diffbar\Psi, \nonumber
\eea
and where  -- for Lie-algebra valued $p$-forms $\omega$ on $\bar{M}$ -- the operator 
adjoint of $\Diffbar$ is
$\Diffbar^\dagger\omega = (-1)^p \astbar\Diffbar\astbar\omega$.
In order to complete the pulsation equations, we still have to
compute the linearized stress-energy tensor:
Using the background equations and
\bdm
\delta\dot{\bar{F}}_{ij} = -\Diffbar_i {\cal E}_j + \Diffbar_j {\cal E}_i, \;\;\;
\delta\dot{ H} = -\Psi, \;\;\;
\hbox{etc...},
\edm
one finds that the expression for $\Lambda_{ij}$ -- defined in
(\ref{Eq-LambdaDef}) with $G_{\mu\nu}$ replaced by
$G_{\mu\nu} - 8\pi G T_{\mu\nu}\, $ -- is given by
\bea
\Lambda_{ij} &=& \Lambda^{V}_{ij}
 - 4G\Trace\left\{ \Diffbar_k\!\left( \alpha\bar{F}_{(i}^{\;\; k}{\cal E}_{j)} \right)
 +\frac{1}{\alpha} {\cal E}_k \Diffbar_{(i}\!\left( \alpha^2 \bar{F}_{j)}^{\; k} \right) \right. \nonumber\\
&-&\left. \gbar_{ij}{\cal E}^l \alpha^k\bar{F}_{kl}
 + \frac{1}{\alpha} \Diffbar_{(i}\!\left( \alpha^2\Diffbar_{j)} H \right)\Psi
 + \gbar_{ij} \alpha^k(\Diffbar_k H)\Psi \right\} \nonumber\\
&+& 4G\alpha\Trace\left\{ \bar{F}^k_{\;\, i} \bar{F}^l_{\; j} L_{kl}
 - \frac{1}{4} \bar{F}_{kl} \bar{F}^{kl} L_{ij} - \frac{1}{8}\gbar_{ij} \bar{F}_{kl} \bar{F}^{kl} L \right\}
\nonumber\\
&-& 4G\alpha \Phi L_{ij} + 2\gbar_{ij} \alpha \Phi L,
\label{Eq-LStressEnergy}
\eea
where $\Lambda^{V}_{ij}$ is defined in Eq. (\ref{Eq-LambdaijDes}).
The matter perturbations arising in $\Lambda_{ij}$
perfectly fit together with the gravitational perturbations
arising in $\Lambda^{(YM)}$ and $\Lambda^{(H)}$, with
exception of the terms involving $\left( L - \frac{A}{\alpha} \right)$.
Hence, one either has to adopt the harmonic gauge, $A = \alpha L$,
for which the operator is symmetric but the
kinetic energy is not definite (see section \ref{Sect-LinADMEq}),
or the maximal slicing condition, $L = 0$, for which the
operator acting on $L_{ij}$ is hyperbolic and symmetric,
but the perturbed lapse is still present.
As in the gravitational case, we adopt the maximal
slicing condition.

\subsection{Summary}

The linearized EYMH equations (with arbitrary compact gauge group)
for a static, purely magnetic background, yield a hyperbolic,
formally self-adjoint operator for the variation of the extrinsic
curvature, the electric one-form and the Higgs momentum.
All these quantities are invariant with respect to both infinitesimal
gauge transformations of the gauge fields and infinitesimal coordinate
transformations within the slices $\Sigma_t$.
The wave operator is constructed from appropriate combinations of the
linearized ADM equations, as defined in Eqs. (\ref{Eq-LambdaDef}) and
(\ref{Eq-LambdaYMHDef}).
In the maximal slicing gauge, the evolution equations are described by the
trace-less part of the tensor $\Lambda_{ij}$, given in 
Eq. (\ref{Eq-LStressEnergy}), as well as the expressions for
$\Lambda^{(YM)}$ and $\Lambda^{(H)}$ defined in Eqs. (\ref{Eq-LEYMHADMMatter1})
and (\ref{Eq-LEYMHADMMatter2}), respectively, where one sets $L = 0$.
The constraint equations are the momentum constraint,
(\ref{Eq-LADMEYMHMomentumCons}) and the Gauss constraint,
(\ref{Eq-LADMEYMHGaussCons}). (Additional constraints involving
also perturbations of the metric, the gauge potential and the Higgs field
themselves are the Hamilton constraint and all evolution equations, which
were differentiated with respect to time in order to construct the
wave operator.)
Furthermore, the trace of the tensor $\Lambda_{ij}$ yields
the following elliptic equation for $A$:
\bea
&& \left( \lapbar - R_{00} \right)A = 2\nabbar^k\left\{ \alpha^l L_{kl}
 + G\Trace( \alpha\bar{F}_{lk}{\cal E}^l) \right\} \nonumber\\
&&\quad + 2G\alpha\Trace\left\{ [ H,\Diffbar^k H]{\cal E}_k + 2V' H \Psi -\bar{F}^{km}\bar{F}^l_{\; m} L_{kl} \right\},
\nonumber
\eea
where the momentum constraint and the background equations
have been used as well.
At least in the pure EYM case, this equation is always solvable,
since either $2R_{00} = G\Trace(\bar{F}_{kl}\bar{F}^{kl}) > 0$
and the operator on the left is invertible or the magnetic
field vanishes and the equation reduces to its vacuum counterpart
(\ref{Eq-WaveAlphaVac}). In the presence of Higgs fields
further investigations are needed, since $R_{00}^{(H)} = -2GV$
is negative in that case (i.e. the strong energy condition does not hold
for the Higgs part of the stress-energy tensor).
In some relevant situations, however, the trace equation causes
no problems: In particular, this equation is void 
for odd-parity perturbations of a spherically symmetric background.

Finally, it is also instructive to compute the differentiated
Hamilton constraint, defined as in (\ref{Eq-Lambda00Def})-- with
$G_{\mu\nu}$ replaced by $G_{\mu\nu} - 8\pi G T_{\mu\nu}$.
This yields $\alpha\Lambda_{00} = -\nabbar^j C_j$, where $C_j$
are the constraint variables defined in Eq. (\ref{Eq-LADMEYMHMomentumCons}).
The system of equations defined by
$(\Lambda_{00}, \hat{\Lambda}_{ij}, \Lambda^{(YM)}_i, \Lambda^{(H)})$
is symmetric in the amplitudes $u \equiv (A, L_{ij}, {\cal E}_i, \Psi)$
with respect to the inner product
\bea
\sprod{ u^{(1)} }{ u^{(2)} } &\equiv& \int\limits_\Sigma \left[
  A^{(1)} A^{(2)} + \gbar^{ik}\gbar^{jl} L^{(1)}_{ij} L^{(2)}_{kl} + \right. \nonumber\\
&+& \left. 2G\Trace\left\{ \gbar^{ij} {\cal E}^{(1)}_i  {\cal E}^{(2)}_j + \Psi^{(1)} \Psi^{(2)} \right\}
\right] \etabar.
\eea
As $\Lambda_{00} = 0$ holds by virtue of the
momentum constraint, this again implies that all terms involving
the amplitude $A$ in the remaining equations $\hat{\Lambda}_{ij}$,
$\Lambda^{(YM)}_i$ and $\Lambda^{(H)}$ are orthogonal to the
(momentum-) constraint manifold. Therefore, the variation of the
lapse does not appear if the evolution equations are projected
on the constraint manifold. The initial value problem is solved 
similarly to the vacuum case, see section \ref{Sect-ADMVacInit}.

\section{The coupling to perfect fluids}

At a first glance, it may be surprising that our formulation also
applies to perfect fluids. It is, however, known \cite{S-Book}
that Lagrangian formulations exist in this case.
As we show in Appendix \ref{App-B}, the linear fluctuations
of a self-gravitating perfect fluid are described by the linearized
metric and the Lagrangian displacement vector.
If the background is static, it will be shown that the wave operator
acts on the components of the extrinsic curvature and the
time-derivative of the displacement vector.

\subsection{The ADM equations}

The stress-energy tensor for a perfect fluid is given by
\bdm
T_{\mu\nu} = (\rho + P) u_{\mu} u_{\nu} + P g_{\mu\nu},
\edm
where $u$ is a time-like vector field, describing the
fluid's motion, normalized such that $\gtens(u,u)=-1$. 
The fields $\rho$ and $P$ denote the energy density and the pressure
of the fluid, respectively.
The ADM split of $u$ is given by
\be
u = \gamma (e_0 + v^i e_i),
\label{Eq-ADMFourVelVec}
\ee
where $\gamma^{-2} = 1 - \gbartens(v,v)$ and $e_0$ is the
previously introduced normal unit vector field orthogonal to $\Sigma_t$.
We choose the sign of $\gamma$ such that the fluid flows in the same
direction as $e_0$.

The relativistic Euler equations are obtained from $\nabla^\nu T_{\mu\nu} = 0$:
\bea
0 &=& u^\mu\nabla^\nu T_{\mu\nu} = (\rho + P) \theta + \nabla_u \rho,
\label{Eq-RelEul1}\\
0 &=& q^\alpha_{\; \mu} \nabla^\nu T_{\alpha\nu} = (\rho + P) a_\mu + q^\alpha_{\; \mu} \nabla_\alpha P,
\label{Eq-RelEul2}
\eea
where $\theta = \nabla_\mu u^\mu$ is the expansion,
$a_\mu = (\nabla_u u)_\mu = u^\nu\nabla_\nu u_\mu$ is the
acceleration vector, and where
$q^\alpha_{\;\beta} = \delta^\alpha_{\;\beta} + u^\alpha u_\beta$
projects onto the spaces orthogonal to $u$.
Inserting the expression (\ref{Eq-ADMFourVelVec}) above yields
\bea
\theta &=& \frac{1}{\alpha} \dot{\gamma} + \gamma K + \frac{1}{\alpha} \nabbar^j (\alpha u_j), \nonumber\\
a_i &=& \frac{\gamma}{\alpha}\left( \dot{u}_i + \gamma\alpha_i \right) + u^j \nabbar_j u_i. \nonumber
\eea
Here, we have assumed that the shift vector $\beta$ vanishes,
such that $u_i = \gamma v^i$. For linear perturbations of a static
background there is no loss of generality in doing so since we will
choose a gauge with $\delta\beta=0$.
The stress-energy tensor gives
\bea
T_{00} &=& \gamma^2 (\rho + P), \nonumber\\
T_{i0} &=& -\gamma (\rho + P)u_i, \label{Eq-TFluid}\\
T_{ij} &=& (\rho + P) u_i u_j + P \gbar_{ij}. \nonumber
\eea
In order to get a closed system of equations, an equation
of state is needed as well. In the following we will only 
assume that this has the form $P = P(\rho)$.

\subsection{Coordinate-invariant quantities}

For a static background one has $G_{i0} = 0$, and it is
consistent to set $u_i = 0$ (i.e. the flow is static).
The ADM equations then reduce to
\bea
0 &=& \frac{1}{\alpha} \lapbar\alpha - 4\pi G (\rho + 3P), \\
0 &=& \bar{R}_{ij} - \frac{1}{\alpha}\nabbar_i\nabbar_j\alpha - 4\pi G \gbar{ij} (\rho - P), \\
0 &=& \nabbar_j P + (\rho + P)\frac{\alpha_j}{\alpha}.
\label{Eq-StaticRelEul}
\eea

Under an infinitesimal coordinate transformation generated
by $(X^\mu) = (f, X^i)$ one has the following behavior:
\bea
\delta u_i &\mapsto& \delta u_i - \alpha f_i, \nonumber\\
\delta\rho &\mapsto& \delta\rho + X^s \nabbar_s \rho, \nonumber\\
\delta P   &\mapsto& \delta P + X^s \nabbar_s P . \nonumber
\eea
Therefore, $\delta u_i$ is a vector-invariant quantity, and using
(\ref{Eq-ADMCoordTrans}), we may also construct the vector-invariant
amplitudes
\bea
\delta\dot{\rho}^{(inv)} &\equiv& \delta\dot{\rho} - \delta\beta^s\nabbar_s\rho, \nonumber\\
\delta\dot{P}^{(inv)} &\equiv& \delta\dot{P} - \delta\beta^s\nabbar_s P. \nonumber
\eea
Using the transformation properties of $\delta u_i$ it is tempting to 
construct full coordinate-invariant amplitudes.
These correspond to Lagrangian deformations of $\gtens$, $\rho$
and $P$ (see Appendix \ref{App-B}). Moreover, it turns out that using the
relations (\ref{BEq-Rel}), one can find a wave equation for
$\Delta\ddot{\gbar}_{ij}$ alone.
Unfortunately, the resulting pulsation equation is not symmetric.

The correct quantities to evolve -- from the philosophy adopted in this 
paper -- are the vector-invariant quantities $\delta K_{ij}$ and $\delta u_i$. 
This is motivated by the fact that for a static background, 
the second relation in (\ref{BEq-Rel}) yields
$\alpha\delta u_i = \dot{\xi}_i$.

\subsection{The linearized equations}

We start with the linearization of the relativistic Euler equations
(\ref{Eq-RelEul1}) and (\ref{Eq-RelEul2}).
As before, we make use of the gauge freedom in order to set $\delta\beta = 0$.
Noting that $\gamma$ is quadratic in $u_i$ and that $u_i$
vanishes on the background, we obtain
\bea
\delta\theta &=& \delta K + \frac{1}{\alpha} \nabbar^j (\alpha \delta u_j), \nonumber\\
\delta a_i &=& \frac{1}{\alpha} \delta\dot{u}_i + \nabbar_i\left( \frac{\delta\alpha}{\alpha} \right).
\nonumber
\eea
Also using the background equation (\ref{Eq-StaticRelEul}), the linearization
of equation (\ref{Eq-RelEul1}) gives
\be
\delta\dot{\rho} = -\frac{1}{\alpha}
\Bigl( \alpha^2(\rho + P)\delta K + \nabbar^k \left[ \alpha^2 (\rho + P) \delta u_k \right] \Bigr).
\label{Eq-DeltaRelEul1}
\ee
Defining the speed of sound $c_s$ by $c_s^2 = \p P/\p\rho$, we
also have
\be
\delta\dot{P} = -\frac{c_s^2}{\alpha}
\Bigl( \alpha^2(\rho + P)\delta K + \nabbar^k \left[ \alpha^2 (\rho + P) \delta u_k \right] \Bigr).
\label{Eq-DeltaRelEul2}
\ee
[The above equations also follow from the relations
(\ref{BEq-Rel}) derived in Appendix \ref{App-B}.]
Finally, using the fact that $P$ is a function of $\rho$ only,
the linearization of equation (\ref{Eq-RelEul2}) yields
\be
\delta( q^\alpha_{\; \mu} \nabla^\nu T_{\alpha\nu}) =
\frac{1}{\alpha} \delta\dot{u}_i + \nabbar_i\left( \frac{\delta\alpha}{\alpha} + \frac{\delta P}{\rho + P} \right).
\label{Eq-DeltaRelEul3}
\ee
An evolution equation for $\delta u_i$ is obtained after
differentiating this equation with respect to $t$
and eliminating $\delta\dot{P}$ using equation (\ref{Eq-DeltaRelEul2}).
Defining $B = \alpha^3 (\rho + P)$, $C = \alpha^2 c_s^2/B$,
\be
W_i \equiv \alpha^2 (\rho + P) \delta u_j
\ee
and recalling that $L = \alpha\delta K$, $A = \delta\dot{\alpha}$, we find
\bea
\p_t \delta( q^\alpha_{\; i} \nabla^\nu T_{\alpha\nu}) &=& \frac{1}{B}\p_t^{\, 2} W_i 
 - \nabbar_i\left( C \nabbar^k W_k \right) \nonumber\\
&+& \nabbar_i \left( \frac{A}{\alpha} - c_s^2 L \right) = 0.
\label{Eq-DeltaEulerPuls}
\eea
Furthermore, equation (\ref{Eq-DeltaRelEul3}) yields the
following circularity condition:
\be
\nabbar_i\left( \frac{\dot{W}_j}{B} \right) - \nabbar_j\left( \frac{\dot{W}_i}{B} \right) = 0.
\label{Eq-FluidCirc}
\ee
This condition would also allow to introduce a potential $\Phi$,
defined by $W_i = B\nabbar_i\Phi$. However, in order to obtain 
symmetric equations, it turns out to be necessary to formulate 
the pulsation equations in terms of $\delta u_i$ rather than $\Phi$.
The condition (\ref{Eq-FluidCirc}) is, therefore, kept as a 
constraint equation for the system. That this constraint propagates
is obvious from Eq. (\ref{Eq-DeltaEulerPuls}).
The circularity condition is used to cast
the evolution equation into a hyperbolic and symmetric form.
Indeed, taking a time derivative and using Eq. (\ref{Eq-FluidCirc}),
one can rewrite Eq. (\ref{Eq-DeltaEulerPuls}) in a manifestly
hyperbolic form without destroying the symmetry:
\bea
0 &=& \frac{1}{B} \p_t^{\, 2}\dot{W}_i - \nabbar^k C \nabbar_k \dot{W}_i
   + F^k \nabbar_i\dot{W}_k - \nabbar^k\left[ F_i\dot{W}_k \right] \nonumber\\
  &+& \frac{1}{B}\nabbar^k( C B_k ) \dot{W}_i + (\nabbar_i\nabbar_k C) \dot{W}^k
   - \frac{C}{B^2} B_i B_k \dot{W}^k \nonumber\\
  &+& C \bar{R}_{ik} \dot{W}^k + \nabbar_i \left( \frac{\dot{A}}{\alpha} - c_s^2\dot{L} \right),
\eea
where $F_k = \nabbar_k(BC) /B$. 

In order to complete the pulsation equations, we have to compute
the linearized stress-energy tensor.
Using (\ref{Eq-TFluid}) and (\ref{Eq-DeltaRelEul2}) in
the expression for $\Lambda_{ij}$ -- defined in
(\ref{Eq-LambdaDef}) with $G_{\mu\nu}$ replaced by
$G_{\mu\nu} - 8\pi G T_{\mu\nu}\, $ -- we find
\bea
\Lambda_{ij} &=& \Lambda^{V}_{ij} - 8\pi G\Bigl( \alpha(\rho-P) L_{ij} \Bigr. \nonumber\\
 &+& \frac{2}{\alpha} \bar{G}_{ijkl} \nabbar^k (\alpha W^l)
  + (1 - c_s^2) \gbar_{ij} \nabbar^k W_k \nonumber\\
 &-& \Bigl. \alpha\gbar_{ij} \left[ c_s^2(\rho + P) - (\rho - P) \right] L \Bigr), 
\label{Eq-LFluidStressEnergy}
\eea
where $\Lambda^{V}_{ij}$ is defined in (\ref{Eq-LambdaijDes}).
In order to obtain symmetric equations, we use the linearized
momentum constraint,
\be
C_i = \alpha^2\delta( G_{i0} - 8\pi G T_{i0} ) = \alpha^2\bar{G}_{ijkl} \nabbar^j\!\left( \frac{L^{kl}}{\alpha} \right)
 + 8\pi G W_i .
\label{Eq-LFluidMomentumCons}
\ee
Defining
\bdm
\Lambda^{(F)} \equiv \p_t \delta( q^\alpha_{\; i} \nabla^\nu T_{\alpha\nu}) + \frac{2}{\alpha} C_i\, ,
\edm
we finally obtain
\bea
\Lambda^{(F)} &=& \frac{1}{B}\p_t^{\, 2} W_i - \nabbar_i\left( C \nabbar^k W_k \right)
+ \frac{16\pi G}{\alpha} W_i \nonumber\\
&+& 2\alpha\bar{G}_{ijkl} \nabbar^j \left( \frac{L^{kl}}{\alpha} \right)
 + \nabbar_i \left((1 - c_s^2) L - L + \frac{A}{\alpha} \right). \nonumber
\eea
This shows that, with the exception of the last term involving
$\left( L - \frac{A}{\alpha} \right)$, the matter perturbations
arising in $\Lambda_{ij}$ perfectly fit together with the gravitational
perturbations arising in $\Lambda^{(F)}$.

Like in the vacuum case, we now adopt the maximal slicing gauge $L=0$.
The evolution equations for $L_{ij}$ and $W_i$ are given by the
trace-less part of $\Lambda_{ij}$ and $\Lambda^{(F)}$,
where one sets $L=0$. The constraint equations comprise the
linearized momentum constraint, (\ref{Eq-LFluidMomentumCons}), and
the circularity condition, (\ref{Eq-FluidCirc}). The trace part
of $\Lambda_{ij}$ yields the following equation for $A$:
\bdm
\left( \lapbar - R_{00} \right)A = 2\nabbar^k (\alpha^l L_{kl}) - 4\pi G (1 + 3c_s^2) \nabbar^k W_k,
\edm
with $R_{00} = 4\pi G (\rho + 3P)$. Here, the momentum constraint has
been used in order to simplify the equation. As long as $\rho + 3P$ is
positive, the operator on the LHS is invertible and thus the equation
for $A$ is solvable.
The differentiated Hamilton constraint -- defined in
Eq. (\ref{Eq-Lambda00Def}) with $G_{\mu\nu}$ replaced by $G_{\mu\nu} - 8\pi G T_{\mu\nu}$ -- yields
\bdm
\Lambda_{00} =
-\frac{1}{\alpha}\nabbar^i \left\{ \alpha^2 \nabbar^j \left( \frac{L_{ij}}{\alpha} \right)
 + 8\pi G W_i \right\}.
\edm
The system of equations defined by
$(\Lambda_{00}, \hat{\Lambda}_{ij}, \Lambda^{(F)})$
is symmetric in the amplitudes $u \equiv (A, L_{ij}, W_i)$
with respect to the inner product
\bea
&& \sprod{ u^{(1)} }{ u^{(2)} } \nonumber\\
&& \equiv \int\limits_\Sigma \left[
  A^{(1)} A^{(2)} + \gbar^{ik}\gbar^{jl} L^{(1)}_{ij} L^{(2)}_{kl}
+ 8\pi G \gbar^{ij} W^{(1)}_i W^{(2)}_j \right] \etabar. \nonumber
\eea
As before, $\Lambda_{00} = 0$ is satisfied by virtue of the
momentum constraint and, as a consequence, the variation of the lapse
does not appear if the wave operator is projected onto the constraint manifold.

Finally, the evolution equations $\hat{\Lambda}_{ij} = 0$ and
$\Lambda^{(F)} = 0$ can be written in a manifestly hyperbolic form by a
further differentiation with respect to $t$, as we have shown above.
The initial value problem is solved similarly to the vacuum case.

\section{Conclusion}

We have shown that the perturbation equations governing linear
fluctuations on a static background can be cast into the form of
a constraint symmetric wave equation for gauge-invariant quantities.
In particular, we have discussed the initial value formulation
for vacuum fluctuations of vacuum space-times, where the adapted
gauge-invariant quantities are the components of the 
linearized extrinsic curvature tensor.
For a spherically symmetric background, the constraint equations
have been eliminated in a natural way, and the equations
of Regge-Wheeler and Zerilli were rederived.
An important new feature of the curvature-based perturbation formalism
presented in this paper is that it admits a natural generalization to 
gravitating matter fields.

As a first example, the pulsation operator governing fluctuations 
on a static, purely magnetic EYMH configuration were derived
for an arbitrary compact gauge group.
These pulsation equations are expected to be valuable in order
to discuss the stability of spherically symmetric solutions
to the EYMH equations with respect to non-spherical symmetric
perturbations. In a forthcoming article \cite{Sarbach}, we will
show that the Bartnik-McKinnon solitons \cite{BK-Soliton} and
the corresponding black holes with hair \cite{HairyBH} admit no
unstable modes in the odd-parity sector with total angular momentum
$\ell\geq 1$.
It should also be interesting to generalize the investigations 
to solutions with Higgs fields or with a negative cosmological
constant \cite{W-Stable}, since some of these solutions are known to
be linearly stable with respect to radial perturbations.
Also, the stability of new static, axially symmetric
configurations \cite{KK-StaticAxi} can be discussed within the 
new framework.

In a second example, we have shown that a symmetric formulation
of the equations governing linear fluctuation of a self-gravitating
perfect fluid exists. Provided that the constraint and the
dynamical variables can be decoupled in a ``symmetric way'',
the resulting equations are expected to be useful in order to study
the stability of static, relativistic stars, or the emission of
gravitational waves from pulsating neutron stars
\cite{M-Fluids}, \cite{AAKS-Fluids}.
In order to discuss the fluctuations of rapidly rotating configurations,
it is our aim to generalize the curvature-based approach to general stationary, 
i.e. non-static, background configurations. However, this requires 
further investigations, since the derivation of the wave operator must
be generalized to cases where the extrinsic curvature of the 
background no longer vanishes.

Finally, we mention that the symmetric formulation could also be
useful for second (or higher) order perturbation theory, since the
$n$'th order perturbation is governed by the same wave operator as
the first oder perturbation, but with a source term depending on
perturbations up to order $n-1$.\\
\\
\noindent
{\bf Acknowledgements}\\
O.S. would like to thank D. Giulini for many helpful discussions.
This work was in parts supported by the Swiss National Science Foundation.

\appendix

\section{Derivation of the Regge-Wheeler and Zerilli equations}
\label{App-A}

In this Appendix, we specialize the general formalism to a static
{\it spherically symmetric\/} background and show how to
project the wave operator onto the momentum constraint
manifold. This will result in a new -- and transparent -- derivation 
of the Regge-Wheeler and the Zerilli equations.

A convenient parametrization of the spherically symmetric
$3$-metric is
\be
\gbartens = \diff x^2 + r^2 \ghattens,
\label{AEq-ADMMetricSph}
\ee
where $x$ is a radial coordinate and where $\ghattens = \diff\Omega^2$
is the standard volume element on $S^2$.
The field $r$ and the lapse $\alpha$ are functions of $x$ only.
The Christoffel symbols with respect to $\gbartens$ become
\bea
&& \chrisbar{x}{xx} = \chrisbar{x}{xB} = 0, \;\;\;
   \chrisbar{x}{AB} = -r r' \ghat_{AB}, \nonumber\\
&& \chrisbar{C}{xx} = 0, \;\;\;
   \chrisbar{C}{xB} = \frac{r'}{r} \delta^C_{\; B}, \;\;\;
   \chrisbar{C}{AB} = \chrishat{C}{AB},
\eea
where capital indices refer to coordinates on the $2$-sphere,
and a prime denotes differentiation with respect to $x$.
The components of the Riemann tensor are given by
\bea
&& \bar{R}^x_{\; AxB} = -r r'' \ghat_{AB}, \;\;\;
   \bar{R}^x_{\; CAB} = 0, \nonumber\\
&& \bar{R}^D_{\; CAB} = 2(1 - r'^2) \delta^D_{\; [A} \ghat_{B]C}. \nonumber
\eea
Einstein's background equations (\ref{Eq-ADMBack}) become
\bea
&& 0 = R_{00} = \frac{(r^2\alpha')'}{r^2\alpha}, \nonumber\\
&& 0 = R_{xx} = -2\frac{r''}{r} - \frac{\alpha''}{\alpha}, \label{AEq-ADMBackSph}\\
&& 0 = R_{AB} = \left( 1 - r'^2 - r r'' - r r'\frac{\alpha'}{\alpha} \right) \ghat_{AB}. \nonumber
\eea
The solutions to these equations are, of course,
the Schwarzschild solutions
\bdm
\alpha^2 = 1 - \frac{2Gm}{r}, \;\;\;
\gbartens = \frac{ \diff r^2 }{\alpha^2} + r^2\diff\Omega^2,
\edm
where $m$ is constant.
Our aim is to write the wave operator $\hat{\Lambda}_{ij}$ in terms of
the background metric (\ref{AEq-ADMMetricSph}). Since the
background is spherically symmetric and invariant under parity
reflection, $\ul{x} \mapsto -\ul{x}$, we can perform a multipole
decomposition in the odd- and the even-parity sectors separately.
As this decomposition can be done in an orthonormal manner,
the resulting wave equations will be symmetric again.

\subsection{The odd-parity sector}

The symmetric tensor field $L_{ij}$ can be expanded in terms of
spherical tensor harmonics. In Appendix D of Ref. \cite{SHB-Odd}
we have shown how these tensor harmonics can be obtained from
the standard scalar spherical harmonics $Y^{\ell m}$.
Since the background is spherically symmetric, perturbations
belonging to different $\ell$ and $m$ decouple; in the
following we shall therefore suppress these indices.

In the odd-parity sector, the expansion is given by
\bea
&& L_{xx} = 0, \nonumber\\
&& L_{xB} = n_1\,u_1\, S_B, \label{AEq-LSecond}\\
&& L_{AB} = n_2\, r u_2\, 2\nabhat_{(A} S_{B)}, \nonumber
\eea
where $S_B = (\asthat\diff Y)_B$ denote the transverse spherical
vector harmonics.
The functions $u_1$ and $u_2$ depend on $x$ only, and the normalization
constants $n_1$ and $n_2$ are chosen such that
\bdm
\sprod{L_{ij}}{L_{ij}} = \int\limits_{0}^{\infty} (u_1^2 + u_2^2) \diff x,
\edm
where $\sprod{.}{.}$ denotes the inner product (\ref{Eq-ADMInnerProduct}).
One finds
\bdm
n_1 = \left[ 2\ell(\ell+1) \right]^{-1/2}, \;\;\;
n_2 = \left[ 2\ell(\ell+1)\lambda \right]^{-1/2},
\edm
where we have defined $\lambda = (\ell - 1)(\ell + 2)$.
Since the perturbation of all scalar quantities vanishes in the
odd-parity case, $u_1$ and $u_2$ are coordinate-invariant and
the perturbation of the lapse vanishes. As a consequence, we obtain a
symmetric wave equation for the amplitudes $u_1$ and $u_2$.
Introducing (\ref{AEq-LSecond}) into the wave equation $\hat{\Lambda}_{ij} = 0$
defined as in (\ref{Eq-WaveLij}) gives
\be
\left( \p_t^{\, 2} - \p_\rho^{\, 2} + \bbS + 2V_{BG} \right) u = 0,
\label{AEq-WaveEqVacSphOdd}
\ee
where
\bdm
\bbS = \left( \begin{array}{ll}
  \frac{r}{\gamma}\left( \frac{\gamma}{r} \right)_{\rho\rho} + \lambda\gamma^2 & 
  2\sqrt{\lambda}\gamma_{\rho} \\ 
  2\sqrt{\lambda}\gamma_{\rho} & \frac{r_{\rho\rho}}{r} + \lambda\gamma^2
\end{array} \right).
\edm
Here, we have defined $\gamma = \alpha/r$ and the new radial coordinate $\rho$ according
to $\diff x = \alpha\diff\rho$. Furthermore,
$V_{BG} \equiv \gamma^2\left( 1 - r'^2 - rr'' - rr'\alpha'/\alpha \right)$
vanishes by virtue of the background equations (\ref{AEq-ADMBackSph}).
The momentum constraint equation yields $0 = C_B = n_1 u_c S_B$, where
\be
u_c = \frac{\gamma}{r} \left( \frac{r}{\gamma}\, u_1 \right)_\rho - \sqrt{\lambda}\gamma\, u_2
\label{AEq-DefConsVar}
\ee
parametrizes the constraint variable.

Our aim is to find a new variable $u_p$, representing the
dynamical variable, such that the wave equation assumes the
form
\be
\left[ \p_t^{\, 2} - \p_\rho^{\, 2} + \left( \begin{array}{cc} V_c & 0 \\ V_{pc} & V_p \end{array} \right) \right] \left( \begin{array}{c} u_c \\ u_p \end{array} \right) = 0.
\label{AEq-WaveConsDynSep}
\ee
On the constraint manifold, $u_c = 0$, one then has
\bdm
\left[ \p_t^{\, 2} - \p_\rho^{\, 2} + V_p \right] u_p = 0,
\edm
which is a symmetric wave equation for the dynamical variable $u_p$.
It turns out that this can be achieved with the ansatz
\bdm
\left( \begin{array}{c} u_c \\ u_p \end{array} \right) =
\bbB\left( \begin{array}{c} u_1 \\ u_2 \end{array} \right), \;\;\;
\bbB = \p_\rho + \left( \begin{array}{cc}
  \frac{\gamma}{r}\left(\frac{r}{\gamma}\right)_\rho & -\sqrt{\lambda}\gamma \\ 
 -\sqrt{\lambda}\gamma & A  \end{array} \right), 
\edm
where the first row of the matrix has been chosen such that
Eq. (\ref{AEq-DefConsVar}) holds, while the second row has been chosen such
that the matrix is symmetric. This guarantees that no first order
derivatives appear in $\bbB^\dagger\bbB$.
The key observation is that the function $A$ can be chosen
such that
\bdm
-\p_\rho^{\, 2} + \bbS = \bbB^\dagger \bbB.
\edm
This is indeed the case if $A = -\frac{r_\rho}{r}$.
As a consequence, the desired wave equation (\ref{AEq-WaveConsDynSep})
is obtained after applying $\bbB$ to the left of the original
wave equation (\ref{AEq-WaveEqVacSphOdd}). This yields
\bdm
\left[ \p_t^{\, 2} + \bbB\bbB^\dagger \right] \left( \begin{array}{c} u_c \\ u_p \end{array} \right) = 0,
\edm
where
\bdm
\bbB\bbB^\dagger = -\p_\rho^{\, 2} + \left( \begin{array}{cc}
 \frac{\gamma}{r}\left( \frac{r}{\gamma} \right)_{\rho\rho} + \gamma^2\lambda & 0 \\
 0 & r\left( \frac{1}{r} \right)_{\rho\rho} + \gamma^2\lambda \end{array} \right).
\edm
Thus, the constraint variables decouple from the dynamical variables,
and the wave equation governing the dynamical degree of freedom is exactly
the Regge-Wheeler equation \cite{RW}.
It is also worthwhile noting that the spatial part of the wave operator can be
factorized as $\bbB^\dagger\bbB$, with $\bbB$ regular. This implies that
the Schwarzschild solution is linearly stable in the odd-parity
sector. It turns out that this factorization can be generalized
in the presence of a SU(2) Yang-Mills field.
This then implies the absence of non-spherically symmetric unstable odd-parity modes
for the Bartnik-McKinnon solitons \cite{BK-Soliton}
and the corresponding black holes with hair \cite{HairyBH} \cite{Sarbach}.

\subsection{The even-parity sector}

In the even-parity sector $L_{ij}$ is expanded according to
\bea
&& L_{xx} = n_1\,\frac{h}{r}\, Y, \nonumber\\
&& L_{xB} = n_2\, q\,\nabhat_B Y, \label{AEq-LEven}\\
&& L_{AB} = -n_3\, h\, e^{(3)}_{AB} + n_4\,g\, e^{(4)}_{AB} \nonumber,
\eea
where
\bdm
e^{(3)}_{AB} = r\ghat_{AB} Y, \;\;\;
e^{(4)}_{AB} = r \left( \nabhat_A\nabhat_B Y + \frac{1}{2}\ell(\ell+1)\ghat_{AB} Y \right).
\edm
The basis is chosen such that $L_{ij}$ is trace-less and
orthonormal with respect to the
inner product (\ref{Eq-ADMInnerProduct}):
\bdm
\sprod{L_{ij}}{L_{ij}} = \int\limits_{0}^{\infty} (h^2 + k^2 + g^2) \diff x.
\edm
One finds
\bdm
n_1 = \sqrt{\frac{2}{3}}, \;\;\;
n_2 = \frac{1}{\sqrt{2\mu^2}}, \;\;\;
n_3 = \frac{1}{\sqrt{6}}, \;\;\;
n_4 = \sqrt{\frac{2}{\mu^2\lambda}},
\edm
where we have defined $\mu^2 = \ell(\ell+1)$
and $\lambda = (\ell-1)(\ell+2)$.

In contrast to the odd-parity case, the
amplitudes $h$, $q$ and $g$ are subject to
residual coordinate transformations of the form
(\ref{Eq-ADMResGaugeTrans}).
Expanding $f = \xi(x) Y$, we find
\bea
n_3\, h &\mapsto& n_3\, h - \alpha^2 r'\xi' + \frac{\mu^2}{2}\frac{\alpha^2}{r}\xi, \nonumber\\
n_2\, q &\mapsto& n_2\, q + \alpha r \left( \frac{\alpha}{r} \xi \right)', \label{AEq-ResGaugeTrans}\\
n_4\, g &\mapsto& n_4\, g + \frac{\alpha^2}{r}\xi, \nonumber
\eea
where $f$ obeys the equation $\nabbar^k (\alpha^2\nabbar_k f) = 0$,
i.e.
\bdm
-\frac{1}{\alpha^2 r^2} \left( \alpha^2 r^2 \xi' \right)' + \frac{\mu^2}{r^2}\xi = 0.
\edm
As the dynamical variables must be coordinate-invariant,
these residual transformations will help us to construct them as
appropriate combinations of the amplitudes $(h,g,q)$.
Using (\ref{AEq-LSecond}) in $\hat{\Lambda}_{ij} = 0$
yields the wave equation
\be
\left( \p_t^{\, 2} - \p_\rho^{\, 2} + \bbS \right) u + \bbb(a) = 0,
\label{AEq-WaveEqVacSphEven}
\ee
where the symmetric matrix $\bbS$ is given by
\be
\bbS = \left( \begin{array}{ccc}  \gamma^2 S_{11} & \sqrt{12}\mu\gamma_\rho & 0 \\
                 \sqrt{12}\mu\gamma_\rho & \gamma^2 S_{22} & 2\sqrt{\lambda}\gamma_\rho \\
                 0 & 2\sqrt{\lambda}\gamma_\rho & \gamma^2 S_{33}  \end{array} \right),
\ee
with
\bea
S_{11} &=& \mu^2 + 6r'^2 - 5rr'' - 9 rr'\frac{\alpha'}{\alpha}, \nonumber\\
S_{22} &=& \mu^2 + 4r'^2 - 4rr'' - 8 rr'\frac{\alpha'}{\alpha} + \frac{r^2}{\alpha^2}(\alpha\alpha')',
\nonumber\\
S_{33} &=& \mu^2 - 2r'^2 - rr'' - rr'\frac{\alpha'}{\alpha}. \nonumber
\eea
The inhomogeneous term $\bbb(a)$, where $a = a(x)$ is the scalar
amplitude parametrizing $A$, $A = \alpha\, a Y$, is found to be
\bea
n_3\, b_1 &=& 2\alpha^2 r' a' - \mu^2\frac{\alpha^2}{r} a \nonumber\\
n_2\, b_2 &=& -2\alpha r \left( \frac{\alpha}{r} a \right)', \label{AEq-BTerm}\\
n_4\, b_3 &=& -\frac{2\alpha^2}{r} a. \nonumber
\eea
Next, the $\Lambda^k_{\; k}$-equation yields
\bdm
\frac{2}{r} \left[ \left( r^2 (\alpha a)' \right)' - \alpha\mu^2 a \right]
 + \frac{4n_3}{\alpha} \left[ \alpha\alpha' \frac{r'}{r} - (\alpha\alpha')' \right] h = 0.
\edm
This equation has already been used in our derivation in
order to eliminate the second derivatives of $a$ in the
expression for $b_1$.
Finally, the linearized momentum constraint yields the
following two equations:
\bea
0 &=& C_x = 2n_3\left[ \frac{\alpha}{r^2}\left( \frac{r^2 h}{\alpha} \right)_\rho
 - \frac{\sqrt{3}\mu}{2}\gamma\, q \right] \frac{Y}{r}, \label{AEq-WaveConsVacSphEven}\\
0 &=& C_B = n_2\left[ \frac{\alpha}{r^2}\left( \frac{r^2 q}{\alpha} \right)_\rho
 - \frac{\sqrt{3}\mu}{3}\gamma\, h - \sqrt{\lambda}\gamma\, g \right] \nabhat_B Y.
\nonumber
\eea

Unfortunately, we did not succeed in finding a transformation
from the original variables $u$ to the dynamical variables $u_p$
in a similarly elegant way like in the odd-parity sector, i.e.
by means of a supersymmetry transformation.
However, the dynamical degrees of freedom are found systematically 
as follows: First, one has three original
variables $u = (h,q,g)$ and two constraint variables, which
are given by Eqs. (\ref{AEq-WaveConsVacSphEven}). We are, therefore,
looking for a single dynamical scalar amplitude.
This amplitude must be invariant with respect to the residual
gauge transformation (\ref{AEq-ResGaugeTrans}).
The simplest way to construct a completely coordinate-invariant
amplitude is to take a linear combination of the
amplitudes $h$, $q$ and $g$. Requiring invariance with respect
to the residual transformation (\ref{AEq-ResGaugeTrans}),
the only possibility (up to rescaling) is
\bdm
Z \equiv n_3\, h + n_2\, r'q - n_4\, f g,
\edm
where the function $f = f(x)$ is given by
\bdm
f(x) = \frac{1}{2}\mu^2 + rr'\frac{\alpha'}{\alpha} - r'^2.
\edm
Taking the corresponding combination of the evolution equations
(\ref{AEq-WaveEqVacSphEven}) yields
\bdm
\left( \p_t^{\, 2} - \p_\rho^{\, 2} \right) Z + \gamma^2\mu^2 Z - 6\alpha\alpha'\frac{r'}{r} Z
 - 6n_2\,\alpha\alpha'\frac{r'}{r}\hat{q} = 0,
\edm
where the background equations (\ref{AEq-ADMBackSph}) have
been used in order to simplify the expression.
The coordinate-invariant amplitude $\hat{q}$ is given by
\bdm
\hat{q} = q - r\frac{n_4}{n_2}\left( \frac{g}{\alpha} \right)_\rho.
\edm
Note that the lapse amplitude $a$ does not appear in the
above equation. This becomes clear when one compares (\ref{AEq-BTerm})
with (\ref{AEq-ResGaugeTrans}): Any gauge-invariant combination
of the evolution equations annihilates the terms involving $a$.
The constraint equations, (\ref{AEq-WaveConsVacSphEven}) yield
\bdm
\left( \frac{Z}{\gamma} \right)_\rho + n_2\, f\hat{q} = 0.
\edm
The amplitude $\hat{q}$ can therefore be eliminated. After
using the background expressions
\bdm
\alpha^2 = r'^2 \equiv N = 1 - \frac{2G m}{r}, \;\;\;
\alpha\alpha'\frac{r'}{r} = \frac{G m N}{r^3},
\edm
we obtain the scalar equation
\bdm
\left( \lambda + \frac{6Gm}{r} \right) \left( \p_t^{\, 2} - \p_\rho^{\, 2} \right) Z
-\frac{12G m}{r^2} N Z_\rho + \mu^2\lambda\frac{N}{r^2} Z = 0.
\edm
Defining $Z = (\lambda + 6Gm/r)\Psi$, this is the
Zerilli equation \cite{Zerilli}.

Finally, we mention that the separation between the constraint
and dynamical variables can also be achieved by a first order
transformation of the form $\Psi = A h + B q + C g + g_\rho$
with suitable functions $A$, $B$ and $C$. This ansatz results
in the Regge-Wheeler equation \cite{US-Diplom}.
A similar result has been obtained in \cite{AAL-CurvPert},
however, in contrast to their derivation, the
procedure adopted here is more natural, since the
requirement of gauge invariance with respect to reparametrization
of the time coordinate implies that only one function must be matched.

\section{Perfect Fluids as Field Theory}
\label{App-B}

In this appendix, we briefly review a field theoretical
formulation of perfect fluids models \cite{S-Book} and
discuss some applications to linear perturbation theory.

The fluid's motion is described by a map
\bdm
F: M \rightarrow \Gamma,
x \mapsto \left( F_1(x), F_2(x), F_3(x) \right),
\edm
where $(M,\gtens)$ denotes space-time and where $\Gamma$ is a
three-dimensional manifold.
The fields $F_1$, $F_2$ and $F_3$ are scalar fields
on $M$ and describe a material coordinate system for
the fluid. We assume that the linear map
$F_{*x}: T_x M \rightarrow T_{F(x)}\Gamma$ has rank
three and that the kernel of $F_{*x}$ is time-like
for all points $x\in M$.

Let $\Omega = \bar{n}(F)\diff F_1 \wedge \diff F_2 \wedge \diff F_3$
be a volume form on $\Gamma$, where $\bar{n}(F)$ describes
the material density. Then, we define a 4-current $J$
by
\bdm
J \equiv \ast( F^*\Omega ),
\edm
where $F^*$ is the pull-back and $\ast$ the Hodge-dual
on $(M,\gtens)$. In local coordinates, we have
\bdm
J_\mu = \bar{n}(F)\eta_{\mu\alpha\beta\gamma} (\nabla^\alpha F_1)(\nabla^\beta F_2)(\nabla^\gamma F_3).
\edm
[As an example, consider $F_a(t, x^a) = x^a - t v^a$, $v^a = \const$
which describes a fluid in constant motion in flat space. Then we have
$J^t = \bar{n}$, $J^i = \bar{n} v^i$.]
The definition of $J$ implies that
\bdm
\ast\diff\ast J = 0,
\edm
thus the particle number is conserved.
Note also that $J$ spans the kernel of $F_*\,$.
The particle density and the 4-velocity, as measured
in the coordinate system $x$, are defined by
\bdm
n = \sqrt{ -\gtens(J,J) }, \;\;\; \hbox{and} \;\;\;
u^\mu = \frac{1}{n} J^\mu,
\edm
respectively.

The dynamics of the fluid is described by a Lagrangian of the form
\be
{\cal L}(F_a, \nabla F_a) = -n u(v).
\label{BEq-FluidLagr}
\ee
Here, $u$ is a function of $v = 1/n$ and describes
the inner energy per particle.
Some calculations show that the stress-energy tensor,
$T^{\mu\nu} = 2\frac{\delta {\cal L}}{\delta g_{\mu\nu}} + g^{\mu\nu} {\cal L}$,
is given by
\bdm
T^{\mu\nu} = (\rho + P) u^\mu u^\nu + g^{\mu\nu} P,
\edm
where $\rho = n u$ is the energy density and
\bdm
P = - \frac{\p u}{\p v}
\edm
is the pressure.
Note that this agrees with the first law of thermodynamics, $\delta u = -P\delta v$,
when the entropy per particle is constant.
Using $T^{\mu\nu} = g^{\mu\nu}{\cal L} -(\p^\mu F_a) \frac{\p {\cal L}}{\p (\p_\nu F_a)}$,
one also finds
\bdm
\nabla_\nu T^\nu_{\; \mu} = (\p_\mu F_a) \left[ \frac{\p {\cal L}}{\p F_a} 
 - \nabla_\nu \frac{\p {\cal L}}{\p (\p_\nu F_a)} \right].
\edm
Hence, the relativistic Euler equations, $\nabla_\nu T^\nu_{\; \mu} = 0$,
are equivalent to the Euler-Lagrange equations with respect to (\ref{BEq-FluidLagr}),
since $F_*$ was required to have full rank.

Consider linear perturbations of a self-gravitating fluid,
described by the (Eulerian) perturbations $\delta g_{\mu\nu}$ and $\delta R_a$.
One can also introduce the Lagrangian perturbation operator $\Delta = \delta + L_\xi$,
where the Lagrangian displacement field $\xi$ is a first order quantity defined
such that the material coordinates do not change under the perturbations:
\be
\Delta F_a = 0, \;\;\; a=1,2,3.
\label{BEq-LagrDef1}
\ee
Since $F_a$ are scalar fields, this means that
\be
\delta F_a = -(\p_\mu F_a) \xi^\mu.
\label{BEq-LagrDef2}
\ee
Therefore, $\xi^\mu$ is uniquely defined up to a vector proportional to $u^\mu$.
Furthermore, under an infinitesimal coordinate transformation generated by
a vector field $X^\mu$, we must have
\be
\xi^\mu \mapsto \xi^\mu - X^\mu
\label{BEq-LagrDef3}
\ee
(modulo a vector proportional to $u^\mu$) in order to maintain
equation (\ref{BEq-LagrDef2}). As an important consequence of (\ref{BEq-LagrDef3}),
Lagrangian perturbations of a tensor quantity are automatically coordinate
invariant to linear order.

Since $\Delta R_a = 0$ and $[\diff, \Delta] = 0$, $\Delta J^\mu$ and therefore
also $\Delta n$, $\Delta u^\mu$, $\Delta\rho$ and $\Delta P$ can be expressed
in terms of $\Delta\gtens$ only.
Using
\bdm
\Delta J_\mu = -\frac{1}{2} (g^{\alpha\beta} \Delta g_{\alpha\beta}) J_\mu + J^\alpha \Delta g_{\alpha\mu}\, ,
\edm
the following relations can be derived in quite an efficient way
\bea
\Delta n &=& -\frac{1}{2} n q^{\alpha\beta} \Delta g_{\alpha\beta}, \nonumber\\
\Delta u^\mu &=& \frac{1}{2} ( u^\alpha u^\beta \Delta g_{\alpha\beta}) u^\mu, \nonumber\\
\Delta\rho &=& -\frac{1}{2} (\rho + P) q^{\alpha\beta} \Delta g_{\alpha\beta}, \label{BEq-Rel}\\
\Delta P &=& -\frac{1}{2} \gamma P q^{\alpha\beta} \Delta g_{\alpha\beta}. \nonumber
\eea
Here, we have also defined the adiabatic index
\bdm
\gamma = \frac{\p\log P}{\p\log n} = \frac{n^3}{P} \frac{\p^2 u}{\p v^2}\, ,
\edm
and $q^{\alpha\beta} = g^{\alpha\beta} + u^\alpha u^\beta$.
These relations, which are well-known (see, e.g., \cite{FI-Stars}),
imply that the Lagrangian perturbations of the particle conservation equation
and the continuity equation are fulfilled. More precisely, we have
\bea
\frac{\Delta}{n} \left( \nabla_u n + n\theta \right) &=&
\nabla_u\!\left( \frac{\Delta n}{n} + \frac{1}{2} q^{\alpha\beta} \Delta g_{\alpha\beta} \right), \nonumber\\
\frac{\Delta}{\rho + P}\left( \nabla_u\rho + (\rho + P)\theta \right) &=&
\nabla_u\!\left( \frac{\Delta\rho}{\rho + P} + \frac{1}{2} q^{\alpha\beta} \Delta g_{\alpha\beta} \right),
\nonumber
\eea
where $\theta = \nabla_\mu u^\mu$.

\end{document}